\documentclass[sigconf]{acmart}

\usepackage{epsfig,endnotes}

\usepackage{url}
\usepackage{comment}
\usepackage{graphicx}
\usepackage{subfig}
\usepackage{balance}
\usepackage{color}

\usepackage{natbib}
\usepackage{verbatim}
\usepackage{algorithm}
\usepackage{algorithmic}

\newcommand{\accuracy}{96.08\%}
\newcommand{\fpr}{0.9\%}

\begin{document}

\copyrightyear{2017} 
\acmYear{2017} 
\setcopyright{acmcopyright}
\acmConference{CCS '17}{October 30-November 3, 2017}{Dallas, TX, USA}\acmPrice{15.00}\acmDOI{10.1145/3133956.3134083}
\acmISBN{978-1-4503-4946-8/17/10}

\date{}

\title{Practical Attacks Against Graph-based Clustering}


\author{Yizheng Chen}
\affiliation{%
  \department{School of Computer Science, College of Computing}
  \institution{Georgia Institute of Technology}
}
\email{yzchen@gatech.edu}

\author{Yacin Nadji}
\affiliation{%
  \department{School of Electrical and Computer Engineering}
  \institution{Georgia Institute of Technology}
}
\email{yacin@gatech.edu}

\author{Athanasios Kountouras}
\affiliation{%
  \department{School of Computer Science, College of Computing}
  \institution{Georgia Institute of Technology}
}
\email{kountouras@gatech.edu}

\author{Fabian Monrose}
\affiliation{
  \department{Department of Computer Science}
  \institution{University of North Carolina at Chapel Hill}
}
\email{fabian@cs.unc.edu}

\author{Roberto Perdisci}
\affiliation{%
  \department{Department of Computer Science}
  \institution{University of Georgia}
}
\email{perdisci@cs.uga.edu}

\author{Manos Antonakakis}
\affiliation{%
  \department{School of Electrical and Computer Engineering}
  \institution{Georgia Institute of Technology}
}
\email{manos@gatech.edu}

\author{Nikolaos Vasiloglou}
\affiliation{%
  \institution{Symantec CAML Group}
}
\email{nikolaos_vasiloglou@symantec.com}


\begin{abstract}

  Graph modeling allows numerous security problems to be tackled in a general
  way, however, little work has been done to understand their ability to
  withstand adversarial attacks. We design and evaluate two novel graph attacks
  against a state-of-the-art network-level, graph-based detection system. Our
  work highlights areas in adversarial machine learning that have not
  yet been addressed, specifically: graph-based clustering techniques,
  and a \emph{global} feature space where realistic attackers without perfect
  knowledge must be accounted for (by the defenders) in order to be practical.
  Even though less informed attackers can evade graph clustering with low cost,
  we show that some practical defenses are possible.

\end{abstract}


\maketitle

\section{Introduction}

Network level detection systems are used widely by the community as the first
line of defense against Internet threats~\cite{pleiades, chen2016financial,
nelms2016towards, lioperational, invernizzi2014nazca, hao2009detecting,
yen2013beehive}. These systems often represent the underlying network traffic
as a graph for various reasons, but most importantly for the computational
efficiency and scalability that graph techniques enable.  These computational
advantages, for example, enable categorical objects (like domain names and IP
addresses) to be transformed into feature vectors in a multi-dimensional euclidean space. This
allows supervised and unsupervised learning to take place with greater
efficiency.

The wealth of new capabilities that statistical learning systems brought to the
security community make them a prime target for adversaries. Several studies
have shown how security systems that employ machine learning techniques can be
attacked~\cite{lowd2005good, wittel2004attacking, rndic2014practical,
xu2016automatically, sivakorn2016robot}, decreasing their overall detection
accuracy. This reduction in accuracy makes it possible for adversaries to evade detection,
rendering defense systems obsolete.

While graph based network detection systems are not immune to adversarial
attack, the community knows little about {\it practical attacks} that can be
mounted against them. As these network detectors face a range of adversaries
(e.g., from script kiddies to nation states), it is important to understand the
adversary's capabilities, resources, and knowledge, as well as the cost they
incur when evading the systems.

In this paper we present the first practical attempt to attack graph based
modeling techniques in the context of network security. Our goal is to
devise generic attacks on graphs and demonstrate their effectiveness against a
real-world system, called Pleiades~\cite{pleiades}. Pleiades is a network
detection system that groups and models unsuccessful DNS resolutions from
malware that employ domain name generation algorithms (DGAs) for their command
and control (C\&C) communications. The system is split into two phases. First,
an unsupervised process detects new DGA families by clustering a graph of
hosts and the domains they query. Second, each newly detected cluster is
classified based on the properties of the generated domains.

To evade graph clustering approaches like Pleiades, we devise two
novel attacks---targeted noise injection and small community---against three
commonly used graph clustering or embedding techniques: i) community discovery,
ii) singular value decomposition (SVD), and iii) node2vec. Using three
different real world datasets (a US telecommunication dataset, a US university
dataset and a threat feed) and after considering three classes of adversaries
(adversaries with minimal, moderate and perfect knowledge) we mount these two
new attacks against the graph modeling component of Pleiades. We show that even
an adversary with minimal knowledge, i.e., knowing only what is available in open
source intelligence feeds and on their infected hosts, can evade detection.

Beyond devising practical attacks, we demonstrate that the attacks are inexpensive
for adversaries. Fortunately, defenders are not without recourse, and detection
systems' parameters can be tuned to be more resistant to evasion. Based on
these discoveries, we make recommendations to improve Pleaides' resilience.

Our paper makes the following contributions:

\paragraph{Two Novel Attacks} 
\label{par:Two Novel Attacks}

The \emph{targeted noise injection attack} improves on prior work that randomly
injects noise; by targeting the injected vertices and edges to copy the graph
structure of the original signal, we force noise into the resulting clusters.
Our \emph{small community attack} abuses the known property of small
communities in graphs to subdivide and separate clusters into one or more
unrelated clusters.


\paragraph{Practical Attacks and Defenses} 
\label{par:Practical Attacks and Defenses}

While more knowledgeable attackers typically fare better, we demonstrate that
even minimal knowledge attackers can be effective: attackers with no knowledge beyond
their infections can render 84\% of clusters too noisy to be useful, and evade
clustering at a rate of 75\%. The above attacks can be performed at low cost to
the adversary by not appearing to be anomalous, nor losing much connectivity.
Simple defenses raise the attacker's costs and force only 0.2\% of clusters to
be too noisy, and drop the success rate to 25\%. State of the art embeddings,
such as node2vec, offer more adversarial resistance than SVD, which is used in
Pleiades.


\section{Background} 
\label{sec:Background}

\subsection{Graph-based Clustering} 
\label{sub:Graph-based Clustering}

Graph clustering is commonly used in security. Community discovery identifies
criminal networks~\cite{nadji2013connected}, connected components track
malvertising campaigns~\cite{chen2017measuring}, spectral clustering on graphs
discovers botnet infrastructure~\cite{pleiades, chen2016financial}, hierarchical clustering
identifies similar malware samples~\cite{bayer2009scalable,
perdisci2010behavioral}, binary download graphs group potential malware
download events~\cite{nelms2016towards, nelms2015webwitness, invernizzi2014nazca}, and newly devised
graph embeddings, like node2vec~\cite{grover2016node2vec}, could further
improve upon the state of the art. Beyond clustering, other graph-based
techniques are used, such as belief
propagation~\cite{rahbarinia2015segugio,chau2011polonium}. Unfortunately, it is
unknown how resistant these techniques are to adversarial evasion.

\subsubsection{Community Detection} 
\label{sub:Community Detection}

There are many ways to detect communities in a graph. Several techniques in
this space rely on a modularity metric to evaluate the quality of partitions,
which measures the density of links inside and outside communities. This allows
an algorithm to optimize modularity to quickly find
communities. The Louvain algorithm~\cite{blondel2008fast} scales to large networks
with hundreds of millions of vertices. Communities are usually
hierarchical~\cite{ravasz2002hierarchical,sales2007extracting,palla2005uncovering};
however, finding sub-communities within communities is a known hard
problem~\cite{braverman2017eth}. This allows attackers to hide sub-communities
in a ``noisy'' community by adding edges.


\subsubsection{Spectral Methods} 
\label{sub:Spectral Methods}

In~\cite{von2007tutorial}, Braverman et al. discuss several popular spectral
clustering strategies. First, a similarity matrix is used to represent the
graph. Each row and each column represent a vertex to be clustered, and the
weight is a similarity score between the corresponding vertices. After proper
normalization, the matrix $M$ is used as input to singular value decomposition
(SVD) of rank $k$, $SVD_k(M) = U\Sigma V^{*} $. When the resulting eigenvectors
(e.g., vectors in $U$) are further normalized, they can be used as an embedding
in a euclidean space for learning tasks. In spectral methods, the
hyperparameter $k$ is usually chosen by first evaluating the scree plot of
eigenvalues to identify the ``elbow'' where higher ranks have diminishing
returns of representing the input matrix.  When the scree plot starts to plateau at the
$i$th eigenvalue, we set $k = i$~\cite{cattell1966scree, wall2003singular}.

Spectral clustering with SVD is known to have limitations when clusters are
imbalanced; this is due to either graphs being scale-free (power law
distribution)~\cite{lang2005fixing}, or when small communities
exist~\cite{kurucz2010geographically}. Unfortunately, both commonly occur in
real-world data. In practice, these small communities are merged into what is
colloquially called the ``death star'' cluster: a large, noisy cluster that
contains many small communities.


\subsubsection{node2vec} 
\label{sub:node2vec}

Contrary to the strong homophily assumption of community detection and spectral
clustering, node2vec~\cite{grover2016node2vec} has the advantage of balancing
homophily and structural equivalence in its embeddings. For example, vertices
that are sink nodes will have similar embeddings. node2vec generates embeddings
of vertices by optimizing the sum of the log likelihood of seeing the network
neighborhood given a vertex $v$, for all vertices on the graph:

\begin{equation} \label{eq1}
\max_{f} \sum{\log{P( N_S(v) | f(v))}}
\end{equation}

Where $f(v)$ is the embedding of vertex $v$, $N_S(v)$ represents the network
neighborhoods of $v$ with a series of vertices obtained by the sampling
strategy $S$. node2vec proposes a sampling strategy by random walks starting
from every vertex on the graph with the following parameters: 1) number of
walks from each vertex, 2) length of each walk, 3) probability to return to the
same vertex (Breadth First Search), and 4) probability to explore out to
further vertices (Depth First Search). Once the walk samples have been
obtained, node2vec uses a tunable \emph{neighborhood size} to get the neighborhood
of vertices. For example, a walk with length 5 $\{v1, v2, v3, v4, v5\}$
generates the following neighborhoods with size 3: $N(v_1) = \{v2, v3, v4\}$,
$N(v_2) = \{v3, v4, v5\}$.

In order to compute the embeddings given $f(v)$, Equation~\ref{eq1} is
factorized as a product of the conditional probability of each vertex in the
neighborhood based on the conditional independence assumption. Each underlying
conditional probability is defined as a sigmoid function, and the embeddings
are learned by stochastic gradient descent (SGD) with negative sampling
optimization. Effectively, node2vec learns embeddings in a fashion similar to
word2vec~\cite{mikolov2013distributed} but does not use skip-grams. Attackers
can target the neighborhood size and sampling parameters to encourage their
vertices to be under-sampled and thus split into multiple noisy clusters.



\subsection{Related Work} 
\label{sub:Related Work}

Existing work in adversarial machine learning has focused on analyzing the
resilience of classifiers. Huang et al.~\cite{huang2011adversarial} categorize
attack influence as either \emph{causative} or \emph{exploratory}, with the former
polluting the training dataset and the latter evading the deployed system
by crafting \emph{adversarial samples}. Following the terminology of Huang et al.,
our work focuses on \emph{exploratory} attacks that target the graph clustering
component of Pleiades. We assume that the clustering
hyperparameters are selected with attack-free labels, and the subsequent
classifier is not polluted when they are trained. Contrary to other
\emph{exploratory} attacks in literature, we face the challenge that the
clustering features cannot be modified or computed directly, and that attackers often
have an incomplete view of the defender's data.

In order to compute optimal graph partitions or vertex embeddings, one needs to
have a \emph{global} view of all objects on the graph. On the contrary, related work can
compute classification features directly from crafting adversarial samples. For
example, features are directly obtained from spam emails~\cite{lowd2005good,
wittel2004attacking}, PDF files~\cite{smutz2012malicious,
rndic2014practical, xu2016automatically}, phishing
pages~\cite{rndic2014practical}, images~\cite{globerson2006nightmare,
papernot2016limitations, sivakorn2016robot, carlini2016towards}, network attack
packets~\cite{fogla2006evading}, and exploits~\cite{wagner2002mimicry,
tan2002undermining}. These security applications classify an object based on
features extracted from only that object and its behavior. This makes the
features of system classifiers more \emph{local}, and enables evasion
techniques such as gradient descent directly in the feature space. We make the
following definition: a \emph{local} feature can be computed from only one
object; whereas a \emph{global} feature needs information from all objects
being clustered or classified.

Since Pleiades uses \emph{global} features, an adversary's knowledge can affect
the success of attacks. For example, if the adversary has full access to the
defender's datasets, she can reliably compute clustering features and is more
equipped to evade than a less knowledgeable attacker. Many
researchers~\cite{tramer2016stealing, papernot2016transferability} have shown
that, even without access to the training dataset, having knowledge about the
features and an oracle to obtain some labels of objects is sufficient for an
attacker to approximate the original classifier.

Biggio et al.~\cite{biggio2013data, biggio2014poisoning} are the first to study
adversarial clustering. They propose a bridge attack, which works by injecting a small number
of adversarial samples to merge clusters. The attackers have perfect knowledge
in their assumption. We distinguish our work by i) considering attackers with
different knowledge levels, ii) evaluating how adversarial graph-clustering in
network security affects the whole system, and iii) quantifying the cost of
attacks. With respect to attack cost analysis, Lowd et al.~\cite{lowd2005adversarial} propose a linear cost function
as a weighted sum of feature value differences for crafting evasive adversarial
samples. Since we do not work directly in the feature space, we propose
different costs for the attacks we present in Section~\ref{sec:Attacks & Threat
Model}.

\emph{
To summarize, our work is novel because we focus on adversarial clustering,
which deals with global features that cannot be directly changed. We also
evaluate capabilities of attackers with various knowledge levels, and quantify
the costs associated with attacks.
}



\section{Threat Model \& Attacks} 
\label{sec:Attacks & Threat Model}

In this section, we describe our threat model and explain our attacks as
modifications to a graph $G$. In practice, the attacker changes the graph based
on the underlying data that are being clustered. For example, if the vertices
in a graph are infected hosts and the domains they query as in Pleiades, the
graph representation can be altered by customized malware that changes its
regular querying behavior.

\subsection{Notation} 
\label{sub:Notation}

An undirected graph $G$ is defined by its sets of vertices (or nodes) $V$ and
edges $E$, where $G = (V, E)$ and $E = \{(v_i, v_j):$ if there exists an edge between
$v_i$ and $v_j, v_i \in V, v_j \in V\}$.
An \emph{undirected bipartite graph} is a
special case where $V$ can be divided into two disjoint sets ($U$ and $V$) such
that every edge connects at a vertex in $U$ and one in $V$, represented as $G =
(U, V, E)$. While the attacks apply in the general case, oftentimes bipartite
graphs appear in security contexts: hosts ($U$) query domains ($V$), clients
connect to servers, malware make system calls, etc. Finally, a \emph{complete}
undirected bipartite graph is where every vertex in $U$ has an edge to every
vertex in $V$.

$\mathcal{G}$ is an undirected graph that represents the underlying data a
defender clusters. The graph clustering subdivides $\mathcal{G}$ into clusters
$C_0, \ldots, C_k$, where $V = C_0 \cup \ldots \cup C_k$. If the graph
clustering method is based on graph partitions, then each cluster $C_i$ is a
subgraph $G_i$, and $\mathcal{G} = G_0 \cup \ldots \cup G_k$. Often when
applied, a defender seeks to cluster vertices either in $U$ or $V$ of the
bipartite graph, for example, cluster end hosts based on the domains they
resolve, or malware based on the system calls they make. An attacker controls an
\emph{attacker graph}, $G \subset \mathcal{G}$.  The adversary uses the targeted
noise injection and the small community attacks described below to change $G$ to
$G'$, by adding or removing nodes and edges from $G$.

{\it These attacks violate the underlying basic assumptions of graph clustering
techniques, which either renders the clustered subgraph $G'$ to be useless to
the defender or prevents $G'$ from being extracted from $\mathcal{G}$ intact
(See Section~\ref{sub:Attacks_Alg}).}


\subsection{Threat Model} 
\label{sub:Threat Model}

Before describing attacker knowledge levels, we discuss knowledge that is
available to all attackers. We assume all attackers have at least one active
infection, or $G \subset \mathcal{G}$. The attacker is capable of using any
information that could be gathered from $G$ to aid in their attacks. We also
assume that an attacker can evaluate clusters like a defender can, e.g., manual
verification. When done with a classifier, an attacker has black-box access to
it or can construct a surrogate that approximates the accuracy and behavior of
the real classifier based on public data. This may seem extreme, but the
plethora of open source intelligence (OSINT)~\cite{cyberwarzone, openiocdb,
iocbucket} data and MLaaS machine learning
tools~\cite{azure,amazon-ml,bigml,predictionio,googleprediction} make this
realistic. Finally, an attacker has full knowledge of the features, machine
learning algorithms, and hyperparameters used in both the unsupervised and
supervised phases of the system under attack, as these are often
published~\cite{pleiades, chen2016financial, nelms2016towards,
invernizzi2014nazca, perdisci2010behavioral, notos, rieck2011automatic}. Since
clustering requires some graph beyond $G$, we must consider attackers with
various representations of the defender's $\mathcal{G}$. We evaluate three
levels: minimal, moderate, and perfect knowledge. The minimal level attacker
only knows what is in their attack graph $G$, but the perfect attacker
possesses $\mathcal{G}$. For example, a perfect adversary would have access to
the telecommunication network data used in Pleiades, which is only obtainable by the
most sophisticated and well resourced of adversaries.

\paragraph{Minimal Knowledge} 
\label{par:Minimal Knowledge}

The minimal knowledge case represents the least sophisticated adversary. In
this case, only the attacker graph $G$ is known, as well as any open source
intelligence (OSINT). For example, the attacker can use OSINT to select
potential data to inject as noise, or can coordinate activities between their
vertices in $G$. In the Pleiades example, an attacker with minimal knowledge
can draw information from their infected hosts.


\paragraph{Moderate Knowledge} 
\label{par:Moderate Knowledge}

The moderate knowledge case represents an adversary with $\mathcal{\tilde{G}}$,
an approximation of $\mathcal{G}$. If attacking Pleiades, $\mathcal{\tilde{G}}$
would be a host/domain graph from a large enterprise or university in order to
approximate the view that the defender has. This allows the adversary to
evaluate their attacks. The size of $\mathcal{\tilde{G}}$ affects the
evaluation from the attacker's perspective, which we will explore by randomly
sampling subgraphs of $\mathcal{\tilde{G}}$. An attacker with moderate
knowledge is similar to a sophisticated adversary with access to large datasets
through legitimate (i.e., commercial data offerings) or illegitimate (i.e.,
security compromises) means.


\paragraph{Perfect Knowledge} 
\label{par:Perfect Knowledge}

Finally, the perfect knowledge case is for an adversary who has obtained
$\mathcal{G}$ from the defender. Given the full dataset and knowledge of the
modeling process, an adversary can completely reconstruct the clustering
results of the defender to evaluate the effectiveness of their attacks.
Ideally, this data would be well guarded making this level of knowledge only
realistic for the most sophisticated of attackers, e.g., nation-state sponsored
threats. Nevertheless, considering the damage that could be done by a perfect
knowledge attacker is important as a security evaluation, since it allows us to find potential
weaknesses in the graph clustering techniques.

\subsection{Attacks} 
\label{sub:Attacks_Alg}

We present two novel attacks against graph clustering. The first,
\emph{targeted noise injection}, improves on random
injections~\cite{sun2005neighborhood, liu2008towards} by emulating the
legitimate signal's graph structure. The second, \emph{small community attack},
exploits the known phenomenon of \emph{small communities} in
graphs~\cite{kurucz2010geographically, li2012small}. Our attacks violate
both the homophily and the structural equivalence assumptions used by graph clustering
methods. That is, our attacks either change what nodes are close together to violate homophily,
or they change observations of node neighborhoods so as to violate structural
equivalence.

Identifying a successful attack depends on the system, which will be described
in detail in Section~\ref{section:evasion}. Since we use Pleiades, we evaluate
attacks by the impact on a subsequent classification of the resulting
adversarial clusters. However, this could be done purely at the unsupervised
level by {\it manually} evaluating the accuracy of the output clusters, or
leveraging prior work in adversarial malware clustering~\cite{biggio2013data}
to measure global cluster quality decrease. Next, we evaluate the cost
incurred by the attacker. We analyze the costs by measuring changes to their graph's structure that
would either flag them as anomalous or damage connectivity between the graph's
vertices. In the descriptions below, an attacker's initial graph $G$ is shown,
and the alterations yield a modified graph, $G'$, that represents a defender's
view of the attacker's graph after the adversarial manipulation.

\subsubsection{Targeted Noise Injection} 
\label{sub:Noise Injection}

\begin{figure}[t]
    \centering
    \scalebox{0.5}{\includegraphics{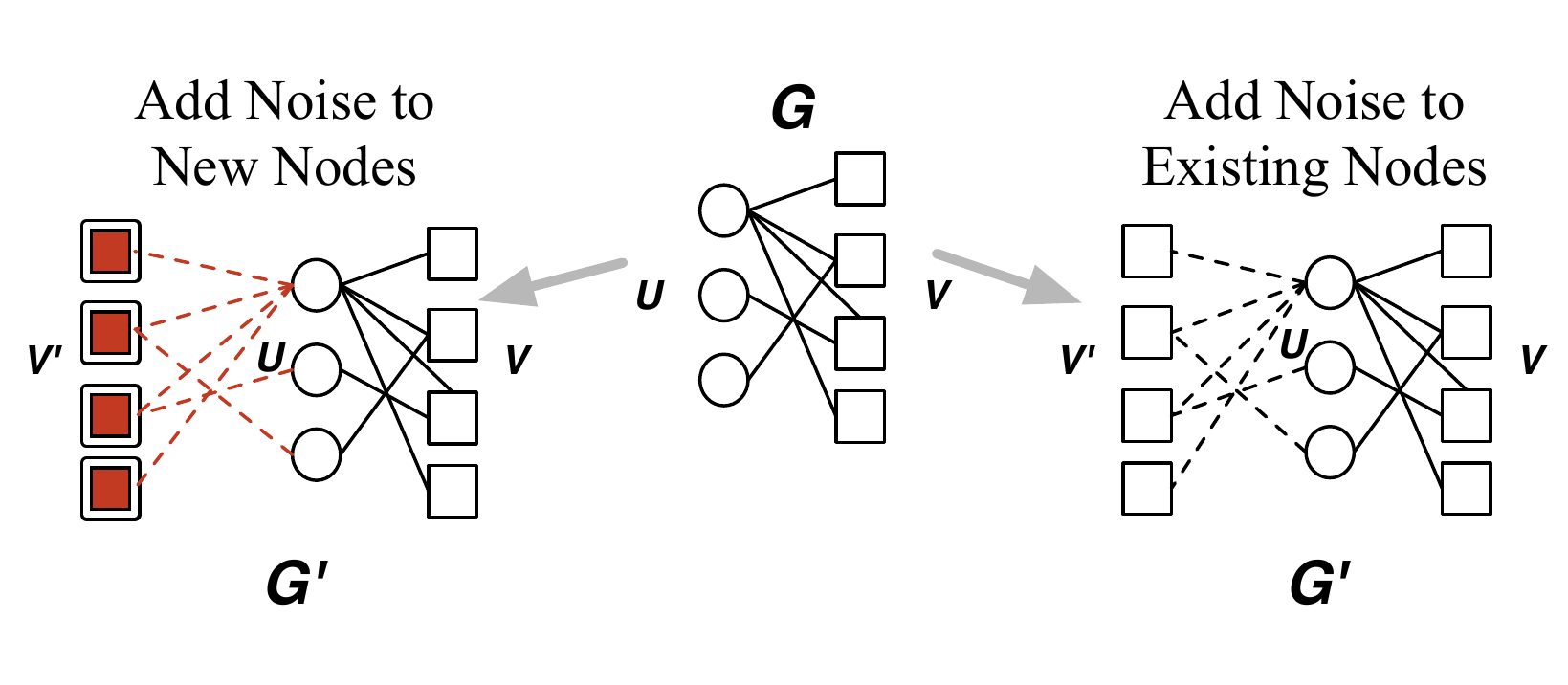}}
    \caption{
    Example of targeted noise injection attacks on a graph.
    }
   \label{fig:noise_injection_example}
\end{figure}

Figure~\ref{fig:noise_injection_example} illustrates two targeted noise
injection attacks. Consider a bipartite attacker graph $G$, with vertex sets
$U$ (circles) and $V$ (squares). To mount the attack, noise is injected into
$G$ to generate $G'$. We inject noisy edges from nodes controlled by the
attacker for the purpose of mirroring real edges. This encourages newly connected nodes to be
clustered together with the attacker's nodes.

To inject noise, the attacker creates an additional vertex set $V'$, represented by red
squares. Entities in $V'$ should differ substantially from those in $V$,
which depend on the underlying system to be evaded. In Pleiades' case, this
means the injected domains ($V'$) must be different, in terms of character
distribution, from the legitimate domains ($V$). Then, for every
edge between $U$ and $V$, the attacker creates a corresponding edge between $U$ and
$V'$, as shown in Figure~\ref{fig:noise_injection_example}. That is, the attack
function $f: (u, v) \in E \mapsto (u, v') \in E'$ is bijective. This creates $G'
= (U, V \cup V', E \cup E')$, where $E'$ are the corresponding edges from $U$
to $V'$, denoted by dotted red edges in the figure. The other way to inject
noise is to create edges from $U$ to existing nodes from $\mathcal{G}$, as
shown in Figure~\ref{fig:noise_injection_example}. This does not add additional
nodes, but identifies other vertices on the defender's graph $\mathcal{G}$ to
use as $V'$. A new edge is created for all edges between $U$ and $V$. Attacker
information is used to identify additional nodes to use. Example nodes may include other
non-malicious domains queried by infected hosts, or a machine's existing
behavior observed by the malware. More commonly, it requires some knowledge of
the graph being clustered, $\mathcal{G}$. This process can be repeated to
increase $|V'|$ to be multiples of $|V|$.

\renewcommand{\algorithmicrequire}{\textbf{Input:}}
\renewcommand{\algorithmicensure}{\textbf{Output:}}

\begin{algorithm}[t]
\caption{Targeted Noise Injection Attack Algorithm for Attacker $\mathcal{A}$ controlling $G$}
\label{alg:noise}
\begin{flushleft}
\algorithmicrequire{ $\mathcal{A}, m, G = (U, V, E)$}\\
\algorithmicensure{ $G'$}
\end{flushleft}
\begin{algorithmic}[1]
	\FOR{$i=1$ \TO $m$}
		\STATE $V'_i \gets$ according to knowledge of $\mathcal{A}$
		\FOR{$v' \in V'_i$}
			\STATE Mirror the edges such that $f: (u, v) \in E \mapsto (u, v') \in E'_i$ is bijective.
		\ENDFOR
	\ENDFOR
	\STATE Return $G' = (U, (\bigcup\limits_{i=1}^{m} V'_i) \cup V, (\bigcup\limits_{i=1}^{m} E'_i) \cup E)$
\end{algorithmic}
\end{algorithm}

Algorithm~\ref{alg:noise} formally describes noise injection for attacker
$\mathcal{A}$ controlling the attacker graph $G$, with noise level $m$.
Line 1 to Line 6 repeats the noise injection process $m$ times.
In line 2, $\mathcal{A}$ generates the set of \emph{noisy nodes} $V'$
according to her knowledge. From line 3 to 5,
the attacker creates a one-to-one mapping from $E$ to $E'_i$.
Line 8 returns the manipulated attacker graph $G' =
(U, (\bigcup\limits_{i=1}^{m} V'_i) \cup V, (\bigcup\limits_{i=1}^{m} E'_i) \cup E)$.
We will evaluate two variants to determine how much noise is needed to mount a
successful, but low cost attack. In the first variant $m=1$, and in the second
$m=2$.

While additional edges and nodes could be injected arbitrarily at random, we
choose to mirror real edges in order to make both nodes from $V'$ and $V$ have
similar embeddings. We define $V'$ to be the set of \emph{noisy nodes}. The
targeted noise injection attack exploits the homophily
assumption~\cite{von2007tutorial, blondel2008fast} of graph clustering methods.
In community discovery and spectral methods, graph partitions cannot
distinguish injected noisy nodes ($V'$) from real nodes ($V$), which exhibit
structurally identical connections to $U$. The co-occurrence increases the
observation of noisy nodes appearing in neighborhoods of real nodes, and vice
versa for node2vec. We expect nodes from $V'$ to join existing clusters
containing $V$.

The targeted noise injection attack has a cost for the attacker of raising the profile of nodes
belonging to attacker graph $G$, potentially making them outliers.
Specifically, hosts from $U$ will increase in percentile with respect to their
degree, i.e., a relatively high degree could indicate anomalous behavior, which
we can measure by the increase in percentile ranking changes before and after
an attack. We call this the \emph{anomaly cost}.

For attacking Pleiades, consider a graph where $U$ are infected hosts and $V$
are the domain names that the hosts in $U$ query. To generate $G'$ an attacker
instructs their malware to query an additional domain ($v \in V'$) for each
domain used for its normal malicious operation ($v \in V$).  This causes the
domains from $V$ and $V'$ to conflict such that the clustering is not
useful to the defender.  However, the anomaly cost may make these trivial to
detect. Nonetheless, we will show in Section~\ref{ssub:Noise Injection
Costs} that the cost of attack is small enough to be practical.


\subsubsection{Small Community} 
\label{sub:Diffusion}

\begin{figure}[t]
    \centering
    \scalebox{0.5}{\includegraphics{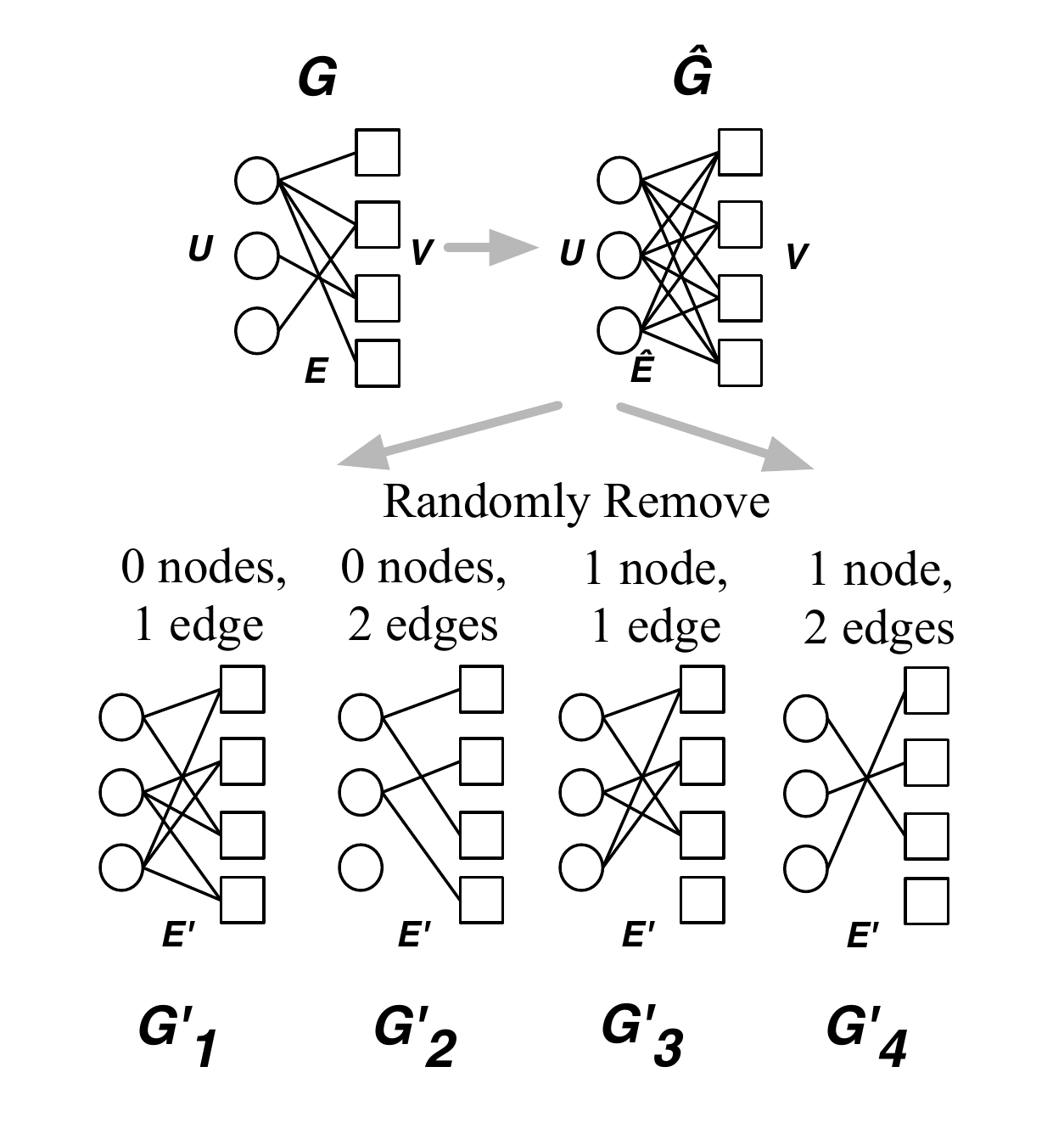}}
    \caption{
    Example small community attacks on a graph.
    }
   \label{fig:diffusion_example}
\end{figure}

Figure~\ref{fig:diffusion_example} illustrates four potential small community
attacks of increasing intensity. The small community attack removes edges
and/or nodes such that the graph clustering separates a single attack graph
into multiple clusters, while maintaining as much connectivity from the
original graph as possible. Again, $G$ is a bipartite attacker graph with
identical vertex and edge sets as before. To mount the attack, an adversary
first constructs a complete version of $G$, $\hat{G}$. In $\hat{G}$,
every vertex in $U$ has an edge to every vertex in $V$. To construct $G'$, the
adversary removes edges from $\hat{G}$. In Figure~\ref{fig:diffusion_example},
the attacker has removed one and two edges per vertex in $V$ in $G'_{1}$ and
$G'_{2}$, respectively. Then in $G'_{3}$ and $G'_{4}$, the attacker has removed
a vertex from $V$, and then removed one and two edges per remaining vertex. The
attacker randomly chooses $n_v$ (such that $0 \leq n_v \leq |V| - 1$) nodes to
remove, and $n_e$ (such that $0 \leq n_e \leq |U| - 1$) edges from each
remaining node $V$ in $\hat{G}$. In the extreme case, there is only one vertex
remaining from $V$ connecting to one in $U$, which often cannot be captured by
graph embeddings. Each attack instance is configured with ($n_v$, $n_e$) pair,
or, in other words, the ($|V|-n_v$, $|U|-n_e$) pair to keep nodes and edges. We
define the \emph{attack success rate} as the number of successful attack
configurations divided by $|U|*|V|$.

If the attacker has minimal knowledge, she can choose $n_v$ and $n_e$
randomly, and hope for the best. With perfect knowledge (knows
$\mathcal{G}$), she can choose the smallest $n_v$ and $n_e$ that succeed.
Attackers without some knowledge or approximation of $\mathcal{G}$ will be
unable to verify if their attacks succeed. While $G$ could be manipulated
directly, removing nodes and edges lowers the utility for the attacker by
losing connectivity in their attack graph. We aim to reduce the average edge
number per node of $V$ in $G$, while simultaneously maintain the lowest possible
cost. Constructing and altering $\hat{G}$ both simplifies the experiments of
quantifying the small community attack cost and makes the job of the attacker
easier.
We believe this does not negate the correctness of our experiments.

\begin{algorithm}[t]
\caption{Small Community Attack Algorithm for Attacker $\mathcal{A}$ controlling $G$}
\label{alg:smallcom}
\begin{flushleft}
\algorithmicrequire{ $\mathcal{A}, G = (U, V, E)$}\\
\algorithmicensure{ $G'$}
\end{flushleft}
\begin{algorithmic}[1]
	\STATE Construct $\hat{G} = (U, V, \hat{E})$ from $G$, where $|\hat{E}| = |U|*|V|$
	\STATE $n_v, n_e \gets$ according to knowledge of $\mathcal{A}$, where $n_v < |V|, n_e < |U|$
	\STATE $V' \gets$ Choose $|V|-n_v$ random nodes from $V$
	\FOR{$v' \in V'$}
		\STATE Choose $|U|-n_e$ random edges that connect to $v'$ to update $U'$ and $E'$
	\ENDFOR
	\STATE Return $G' = (U', V', E')$
\end{algorithmic}
\end{algorithm}

Algorithm~\ref{alg:smallcom} formally shows the the small community attack
$\mathcal{A}$ in control of graph $G$. Line 1 constructs the complete graph
$\hat{G}$ from $G$.  In line 2, $\mathcal{A}$ chooses $(n_v, n_e)$ according to
her knowledge. Then, the attacker chooses $|V|-n_v$ random nodes from $V$ as
$V'$. Each node in $V'$ connects to all nodes in $U$. From line 4 to 6, the
attacker chooses $|U|-n_e$ random edges to keep for each node in $V'$, and thus
forming $U'$ and $E'$. Lastly, line 7 returns $G' = (U', V', E')$ as the
manipulated attacker graph.

The small community attack exploits the information loss in graph embeddings.
While community discovery works better at identifying islands and singletons,
graph embeddings may miss such signal given the hyperparamters chosen at
deployment. Existing methods for choosing hyperparameters do not account for
potential small community attack opportunities. The downside of the attack is
the \emph{agility cost}. By removing nodes and edges from $G$, the adversary
has to give up control over nodes, redundancy, or even functionality. In
addition to losing $n_v$, the agility cost can be measured by the change in
\emph{graph density} (Equation~\ref{eqn:density}) from $\mathcal{D}(G)$ to the
chosen $\mathcal{D}(G')$. We define the following $D(G)$ and $D(G')$:

\begin{equation}
  \label{eqn:density}
  \mathcal{D}(G) = \frac{|E|}{(|U|*|V|)}
\end{equation}

\begin{equation}
  \label{eqn:attackdensity}
  \mathcal{D}(G') = \frac{|E'|}{(|U|*|V|)}
\end{equation}

A graph's density ranges from [0, 1], which denote a graph with zero edges or
all possible edges, respectively. For $D(G')$ we consider how many edges are in
$E'$ compared to the maximum possible number of edges between the original $U$
and $V$. This normalizes the number of edges by the structure of the $G$. The
\emph{agility cost} is $D(G) - D(G')$ if $D(G) > D(G')$, or zero if $D(G) \leq
D(G')$. A loss in density implies a potential loss of connectivity, but
maintaining or increasing the density bodes well for the attacker.  They can
afford an even denser structure, yet still evade defenders. It is important to
note that, while $n_v$ is lost, this is reflected in the density score, as $|V|$
includes any removed vertices like $n_v$. A lower density, and therefore a
higher cost, is incurred when edges and/or vertices are removed relative to the
original structure seen in $G$.

Consider an attack on Pleiades. $\hat{G}$ is created by completing $G$. To
mount the attack like $G'_{2}$, the adversary partitions the domain names that
are used to control her malware by removing one of the control domains
($n_v$=1), and then excludes two distinct hosts that query each of the remaining
domains ($n_e$=2). In other words, the adversary can also randomly choose one
host ($|U| - n_e$) to query each one of remaining control domains. This reduces
the density from $D(G)=0.5$ to $D(G'_{2}) = 1/3$, and sacrifices one node
($n_v$).

If the adversary has knowledge that allows testing whether the attack is
successful or not, the attacker can increasingly remove domains and queries
from hosts until clustering $\mathcal{G}$ no longer results in $G'$ being
extracted as a single cluster. In practice, as described in
Section~\ref{sub:Graph-based Clustering}, the subdivided $G'$ often ends up
either as portions of the ``death star'' cluster; or in multiple, noisy clusters.
In both cases, the legitimate cluster is effectively hidden in a forest of noise. In order to
verify an attack was successful, however, an attacker must have $\mathcal{G}$
or an approximation.




\section{Attacks in Practice}
\label{section:evasion}

We chose to attack Pleiades because it has been commercially deployed and
relies on graph modeling. Our reimplementation has similar performance, as
shown in Appendix~\ref{section:pleiades}. We now describe portions of the
reimplementation in detail.

\subsection{Pleiades} 
\label{sub:Pleiades}

\begin{figure}[t]
    \centering
    \scalebox{0.4}{\includegraphics{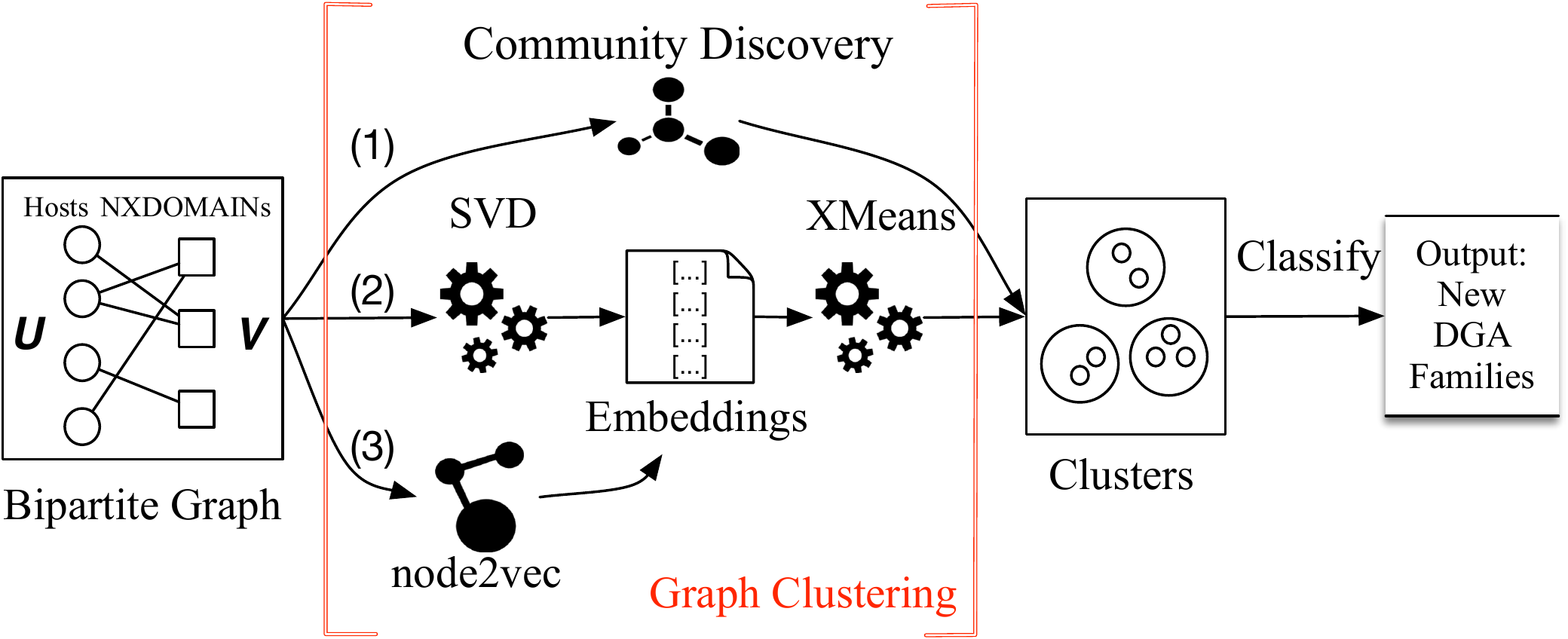}}
    \caption{
    Overview of the DGA detection system.}
   \label{fig:detector}
\end{figure}

An overview of Pleiades is shown in Figure~\ref{fig:detector}. We focus our
attacks on the clustering component and use the classification phase to
demonstrate attack success. First, Pleiades clusters NXDOMAINs ($V$) queried by
local hosts ($U$) using the host-domain bipartite graph ($\mathcal{G}$). It
groups together NXDOMAINs queried by common hosts into clusters $C_0, \ldots,
C_k$, based on the assumption that hosts infected with the same DGA malware
query similar groups of DGA domains. The graph clustering can be achieved by
either community discovery (1), spectral clustering (2) as in the original
paper~\cite{pleiades}, or node2vec clustering (3). Then the classification
module computes domain name character distributions of each cluster into a
numerical feature vector, which is used to classify known DGA families. A new
unknown DGA family with features statistically similar to a known one can be
detected by this process. The system operates on daily NXDOMAIN traffic
generated by all hosts in a network using the following data sources.

\subsubsection{Datasets} 
\label{ssub:Datasets}

\begin{table*}[t!]
  \begin{center}
    \begin{tabular}{lrccc}
      \textbf{Dataset} & \textbf{Number of Records} & \textbf{Minimal} & \textbf{Moderate} & \textbf{Perfect} \\
      \hline
      Reverse Engineered DGA Domains     & 14 DGA Families;  395 thousand NXD & X & X & X  \\
      Host-NXDOMAIN Graph (Surrogate)    & 8782 hosts; 210 thousand NXD & - & X & X  \\
      Host-NXDOMAIN Graph (Ground Truth) & average 262 thousand hosts; 1.8 million NXD & - & - & X  \\
      \end{tabular}
    \caption{Summary of datasets and their availability to minimal, moderate, and
    perfect knowledge attackers.}
    \label{tab:dataset-summary}
  \end{center}
\end{table*}

We use anonymized recursive DNS traffic from a large telecommunication company
from December 18, 2016 to December 29, 2016. The dataset contains NXDOMAINs
queried by hosts and the query timestamps. On average, there are 262 thousand
unique anonymized hosts, with 44.6 million queries to 1.8 million unique
NXDOMAINs in a day. We use this dataset to construct Host-NXDOMAIN Graph as
ground truth without attack. This is available to defenders and perfect
knowledge attackers.

As a surrogate network dataset, we use NXDOMAIN traffic from a large US
university network collected on December 25, 2016. It contains 8,782 hosts and
210 thousand unique NXDOMAINs. Among these NXDOMAINs, only 227 appeared in the
ground truth network dataset. The surrogate dataset is available to attackers
with moderate and perfect knowledge.

Last but not least, we use a reverse engineered DGA domains dataset to train
the supervised component of the system. We run the reverse-engineered
algorithms~\cite{github_dga} to generate DGA domains for 14 malware families:
Chinad, Corebot, Gozi, Locky, Murofet, Necurs, NewGOZ, PadCrypt ransomware,
Pykspa, Qadars, Qakbot, Ranbyus, Sisron, and Symmi. The training dataset also
includes live DGA domains observed in the ground truth network. We label
267 clusters belonging to four malware families present in the ground truth
network dataset (Pykspa, Suppobox, Murofet, and Gimemo), and manually verify
that these subgraphs are attack free. We train a Random Forest classifier with
an average accuracy of \accuracy{}, and a false positive rate of \fpr{}. The
classifier trained from this dataset is available for attackers of all
knowledge levels. Table~\ref{tab:dataset-summary} summarizes these datasets.

We discovered 12 new DGA malware families in only 12 days using the ground
truth network traffic (see Appendix~\ref{section:pleiades} for details).  We
discovered real but unsuccessful evasion attempts in the wild, and retrained
our classifier with evasive instances. We believe we have faithfully reimplemented
Pleiades because we use comparable datasets and we achieve similar clustering
and modeling results.



\subsection{Attacks} 
\label{sub:Attacks}

Using the notation described in Section~\ref{sec:Attacks & Threat Model}, let
$\mathcal{G}$ be a bipartite graph of the defender. $U$ represents hosts, both
infected and uninfected, and $V$ represent NXDOMAINs queried by hosts in the
underlying network. An edge exists from $v_i \in U$ and $v_j \in V$ iff the
$i$th host queried the $j$th NXDOMAIN. For an attacker graph $G \subset
\mathcal{G}$, the hosts in $U$ are infected hosts under the control of the
attacker. In the noise injection case, the attacker instructs their malware to
query NXDOMAINs beyond what is needed for normal operation, as shown in
Figure~\ref{fig:noise_injection_example}. In the small community case, the attacker
coordinates the querying behavior of their malware such that they query fewer
NXDOMAINs in common, as in Figure~\ref{fig:diffusion_example}. We will
evaluate the effectiveness of the attacks by the drop in predicted class
probabilities and the predicted label of the classifier. In a Random Forest,
the predicted class probabilities of a feature vector are calculated as the
average predicted class probabilities of all trees in the forest. In practice,
if the predicted class probability decreases substantially, the classifier will
incorrectly label the instances, and the attack will be considered successful.


\subsection{Attack Costs} 
\label{sub:Attack Costs}

To compute the anomaly cost for noise injection, we analyze percentile changes
of edges related to hosts in $U$ in the structure of $\mathcal{G}$ from before
and after the attack. We quantify this change by computing the cumulative
distribution functions (CDFs, example in Appendix~\ref{sub:cdf}) of vertex
degrees before and after a successful attack is mounted. Concretely, if an
attacker can evade Pleiades but raises the profile of their infected hosts from
the $50^{th}$ (in the CDF before attack) to the $99.9^{th}$ percentile of
NXDOMAINs queried per host (in the CDF after attack), a defender will be able
to detect such behavior with simple thresholding (i.e., monitoring hosts
entering the $95^{th}$ percentile).

To quantify the adversarial cost behind the small community attack, we measure
the change of attacker graph density $D(G')$ as defined in
Section~\ref{sub:Diffusion}. If the attacker graph density decreases, this
means the attacker no longer uses NXDOMAINs for their infection and/or the
infected hosts query fewer NXDOMAINs in common, reducing their connectivity
overall and increasing the botnet's management cost.


\section{Results} 
\label{sec:Results}

First, we show how to select hyperparameters for each of the three graph
methods. Next, we present our results for both attacks against each graph based
clustering technique, for the three knowledge levels. Finally, we explain the
costs incurred by the attacker, and how these can be used to identify possible
defenses.

\paragraph{Summary of Results} 
\label{par:Summary of Results}

Our attacks against the graph clustering component of Pleiades gravely reduce
the predicted class probability of the subsequent classification phase. Even
with minimal knowledge, an adversary can launch an effective targeted noise
injection attack dropping the median predicted class probability to 0\%.
In the higher knowledge levels, the maximum predicted class probability can
be as low as 10\%. Using a set of labeled DGA malware families observed in
spectral clustering, the attacks reduce the prediction accuracy from 99.6\%
to 0\%.

In addition to being effective, the attacks do not substantially raise the
anomaly profile of infected hosts: before and after the targeted noise injection
attacks the hosts occupy a similar percentile for the number of NXDOMAINs
queried. Small community attack results show that the traditional way of choosing
hyperparameters for generating graph embeddings is insufficient when we analyze
the system in an adversarial setting, because it creates a large area for
possible small community attack instances. While following the accepted methodology
for selecting the rank for SVD and hyperparameters for node2vec, all DGA
clusters can be hidden in noisy clusters by subdividing the infected hosts into
smaller groups to sacrifice some agility, even while using hundreds of DGA
domains. Even in the minimal knowledge case where the small community attack
cannot be tested, attackers can sometimes still hide.


\subsection{Choosing Hyperparamters}
\label{sub:parameters}

First, we carefully choose hyperparameters for the graph clustering methods in
Pleiades to ensure high quality clusters are generated.
Figure~\ref{fig:parameters} shows the results used to determine the appropriate
hyperparameters.

\begin{figure}[t!]
    \centering
    \scalebox{0.35}{\includegraphics{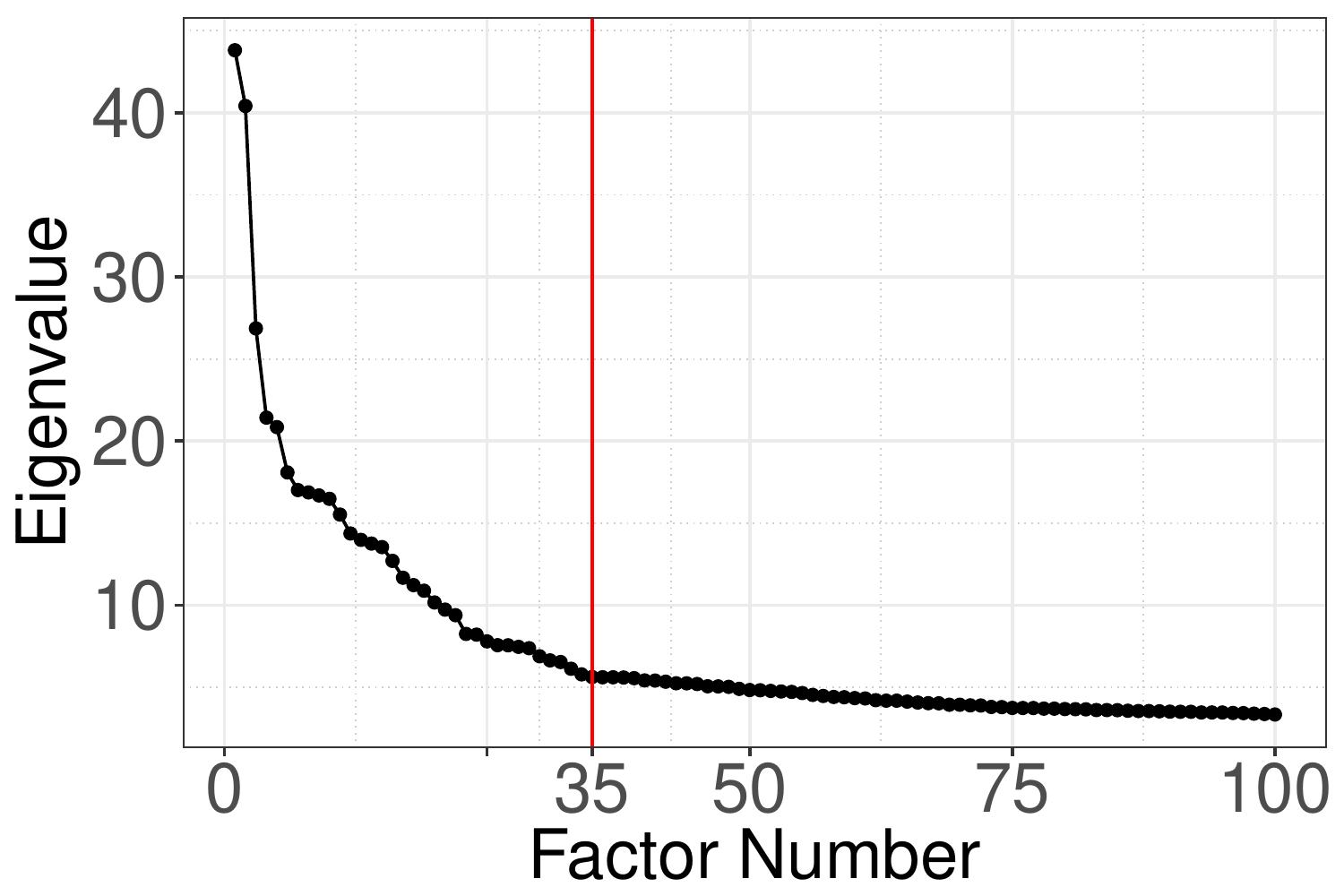}}
    \caption{
    Scree plot of eigenvalues of SVD.}
   \label{fig:screeplot}
\end{figure}

\subsubsection{Spectral Clustering}

We use the scree plot shown in Figure~\ref{fig:screeplot} to choose the rank of
SVD. We fix the rank of the SVD to be 35, where the scree plot plateaus.
While different than the 15 used in the original Pleiades
implementation~\cite{pleiades}, the underlying datasets are different so it is not
unreasonable to find different ranks.

\subsubsection{Community Discovery}

We use the best partition method from the NetworkX community discovery
library~\cite{networkx} which implements the Louvain
algorithm~\cite{blondel2008fast}. The Louvain algorithm efficiently
extracts good communities on the graph by optimizing the modularity metric,
which measures the density of links within communities in comparison to outside
them. It first sets each node to be its own community, and iteratively merges
them to maximize modularity. The algorithm stops when a local maxima is
reached. This community detection algorithm scales to large network with
hundreds of millions of nodes.

\subsubsection{node2vec}
\label{section:node2vec_validity}

\begin{figure}[t!]
    \centering
    \scalebox{0.38}{\includegraphics{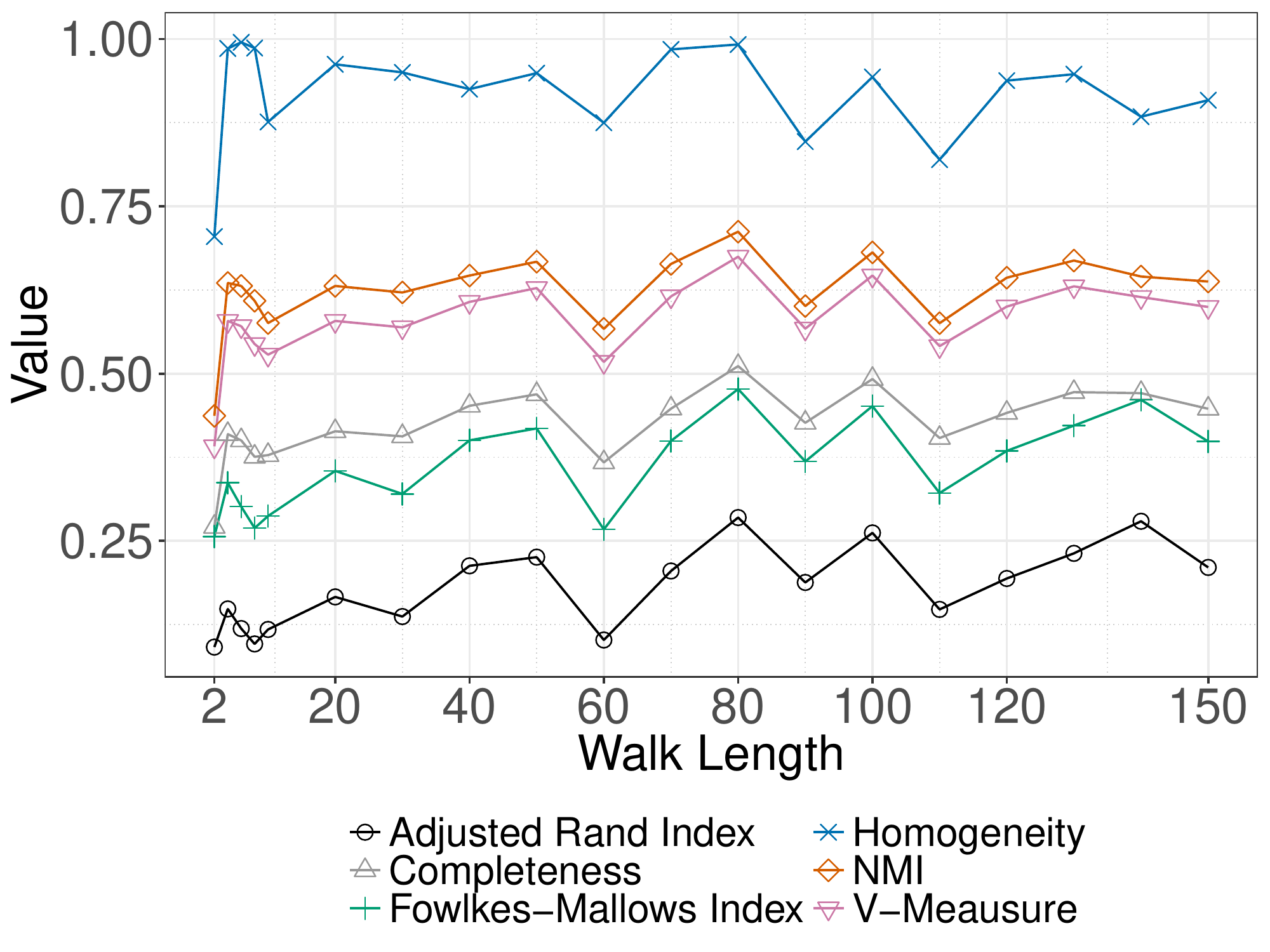}}
    \caption{
    Using cluster validity metrics to choose walk length.}
   \label{fig:validity_walklength}
\end{figure}

We use traditional cluster validity metrics to compare different
hyperparameters of node2vec. Twelve DGA malware families, including both known
and newly detected ones, were used as \emph{reference clusters}. We use validity
metrics including Adjusted Rand Index, Completeness, Fowlkes-Mallows index,
Homogeneity, Normalized Mutual Information (NMI), purity, and V-Measure score.
We first choose context size six, which has the first highest validity scores.
Several larger context sizes generate equal validity scores, but they produce
noisier clusters. This is because that larger context sizes in DNS graphs tend to include
more noisy nodes, such as popular benign NXDOMAINs, or popular hosts that are
likely proxies.

Then we choose the walk length according to
Figure~\ref{fig:validity_walklength}. Multiple walk length values produce high
validity scores, but we choose walk length 20, which corresponds to the second
highest peak. Because using walk length 20 generates cleaner Murofet clusters than a walk length
smaller than 10, due to the fact that longer walk length provides more samples
of neighborhoods and the model is learned better. The number of walks per node,
dimensions, and SGD epoch does not show much difference. We decide on 15 walks,
60 dimensions, and one learning epoch after manual inspection. Lastly,
we use a uniform random probability to choose the next node in the
random walk process.

\subsection{Targeted Noise Injection} 
\label{sub:Noise Injection}

We run our version of Pleiades to generate all attacker graphs. Four DGA
families were identified: Pykspa, Suppobox, Murofet, and Gimemo. For each we
extract the attacker graphs ($G$) and the \emph{\textbf{target domains}} ($V$).
These domains are labeled using the classifier from
Section~\ref{ssub:Datasets}. Before and after the attack, there can be multiple
clusters formed within $G$ and $G'$, depending on the graph clustering
technique. We use the classifier model to test how likely it is that each cluster belongs
to the true DGA malware family, both before and after the attack. We present the
overall distribution of the \emph{predicted class probabilities} to show the
impact of the attacks.

We use different types of noisy domains at different knowledge levels. For a
DGA, these nodes are new NXDOMAINs ($V'$) that will be classified as
benign, also queried by the infected hosts ($U$). In the minimal knowledge
case, we create a DGA algorithm that is classified as benign. It is a
dictionary DGA that uses the most popular English words from movie
subtitles~\cite{subtitles}, popular web terms, and the top one million Alexa
domains. We randomly add numbers and dashes, and randomly select a top-level
domain from four choices. In addition, we generate some punycode domains that
start with the character sequence ``xn--'', and some domains with a ``www''
subdomain. We generate 59,730 verified NXDOMAINs. In the perfect and moderate
knowledge cases, the adversary uses existing, unpopular NXDOMAINs from
$\mathcal{G}$ and the surrogate dataset, respectively.

\subsubsection{Spectral Clustering}

\begin{figure*}[t!]
	\centering
        \subfloat[Spectral Clustering: Predicted class probabilities.]
        {\includegraphics[scale=0.3]{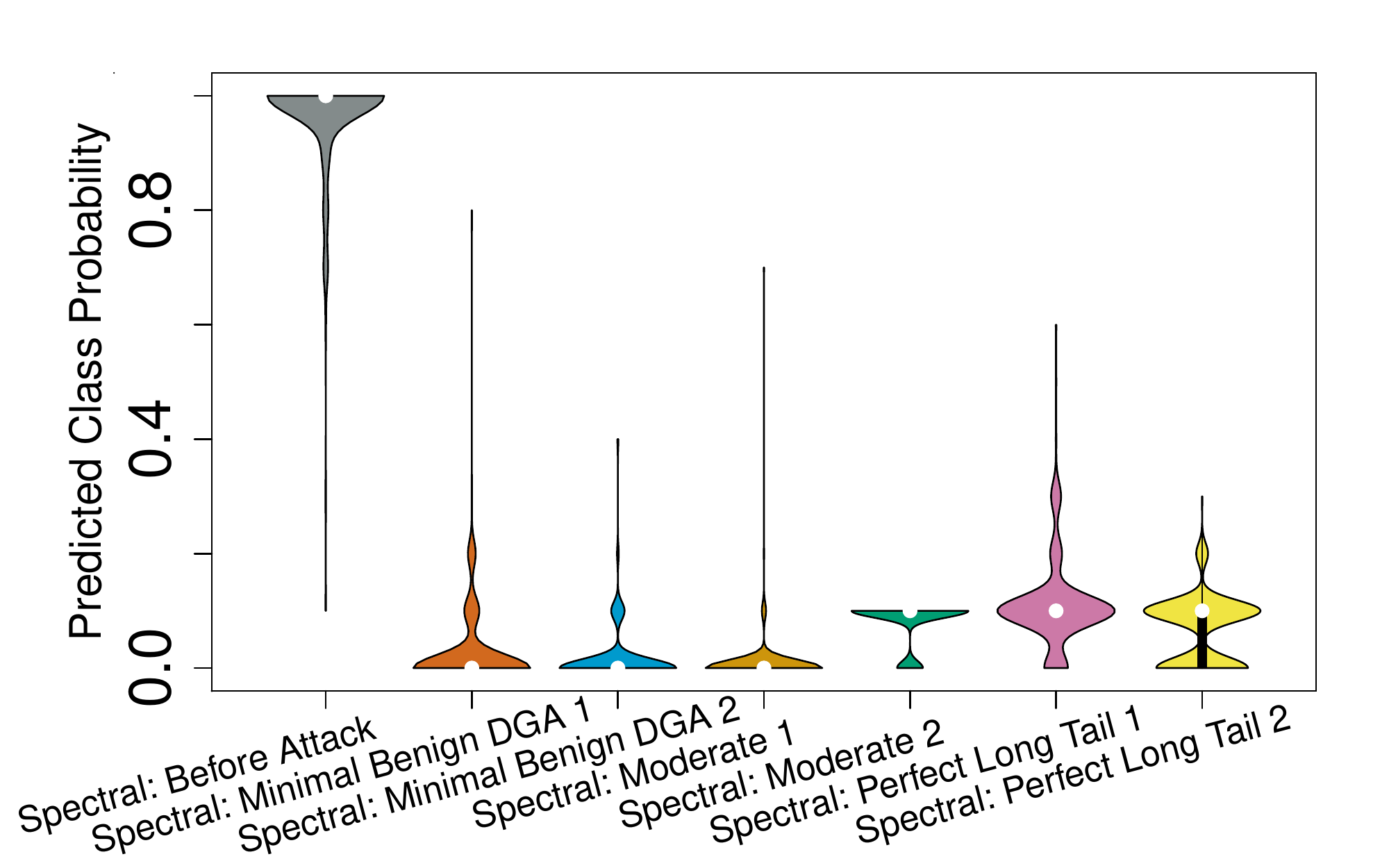}
        \label{fig:prediction_confidence}}
	~
        \subfloat[Community Discovery and node2vec: Predicted class probabilities.]
        {\includegraphics[scale=0.3]{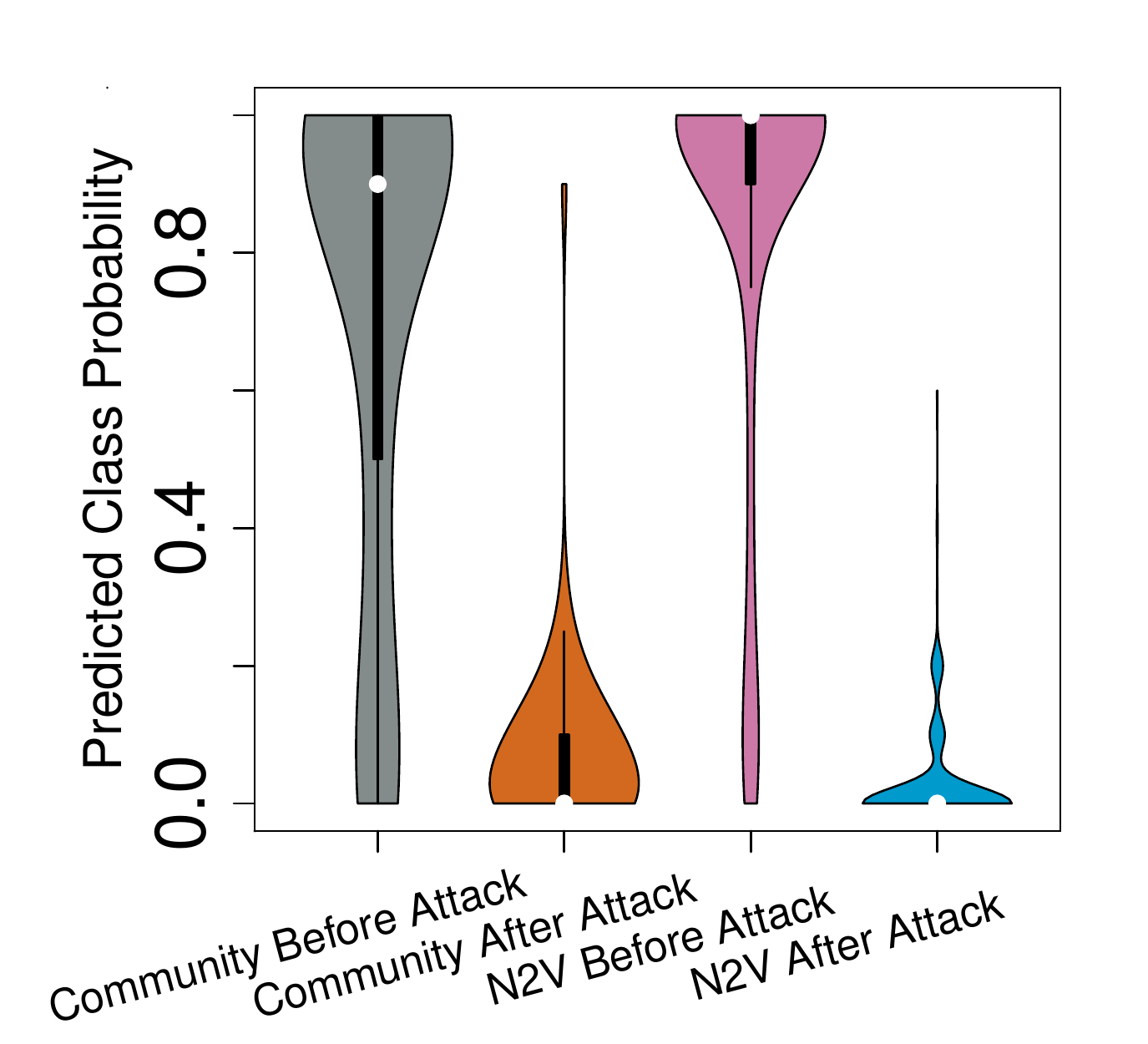}
        \label{fig:prediction_confidence_comm_n2v}}
	~
        \subfloat[Retraining: Predicted class probabilities.]
        {\includegraphics[scale=0.3]{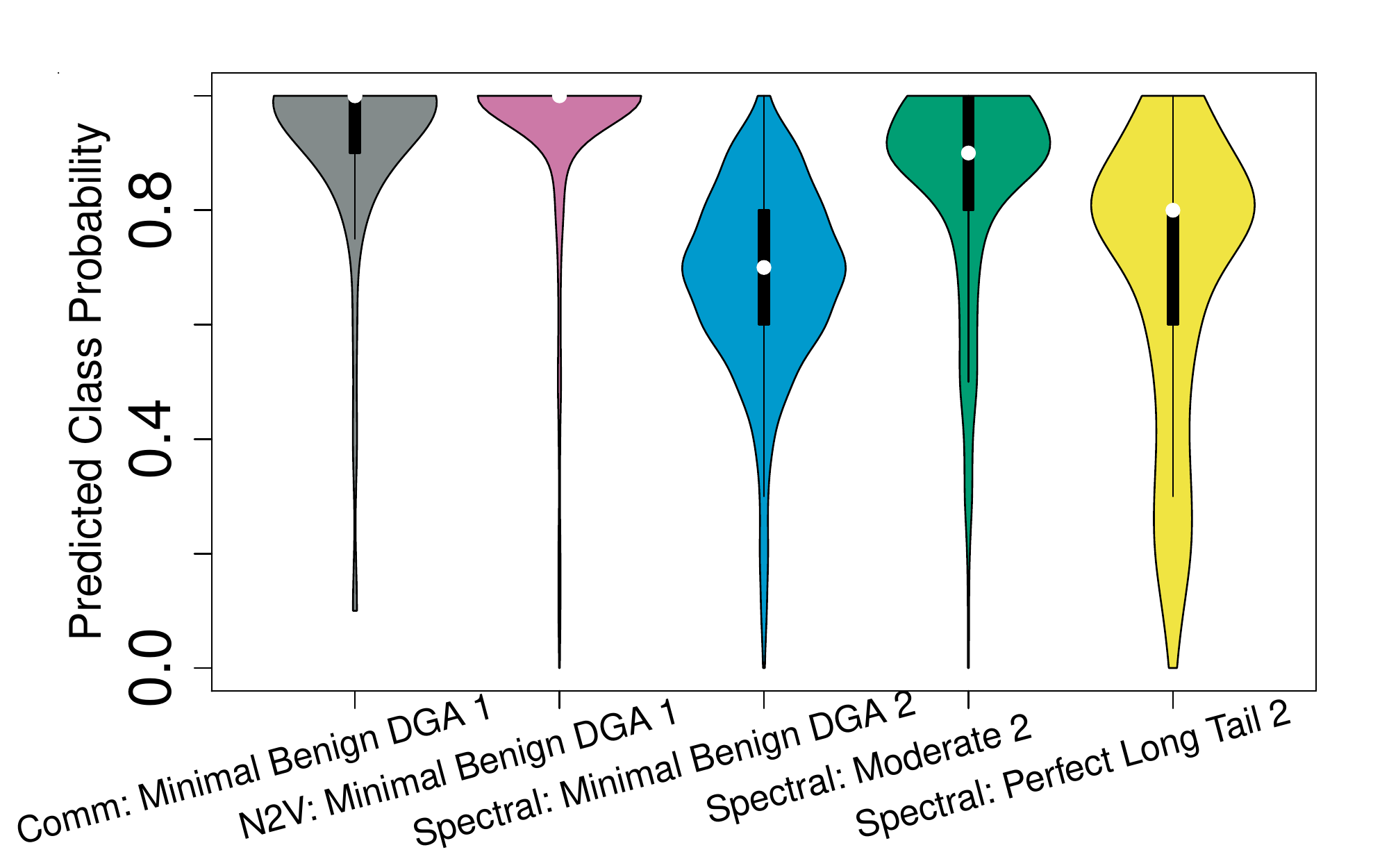}
        \label{fig:retrain}}

        \caption{
         Figure~\ref{fig:prediction_confidence}: Predicted class probabilities
         before the targeted noise injection attack
    	and after two variants of the targeted noise injection attack in minimal, moderate,
	and perfect knowledge.
	Figure~\ref{fig:prediction_confidence_comm_n2v}:
        Predicted class probabilities
        before and after the targeted noise injection attacks
        for community discovery and node2vec.
        Figure~\ref{fig:retrain}: Predicted class probabilities under different attacks
        after retraining including the ``Minimal Benign DGA 1'' clusters.
        }
\end{figure*}

Figure~\ref{fig:prediction_confidence} shows the classifier's predicted true
class probabilities from before the attack is mounted, and after the minimal,
moderate, and perfect knowledge targeted noise injection attacks are performed.
For each knowledge level, we inject two different levels of noise as described
in Section~\ref{sub:Noise Injection} and re-run the clustering and subsequent
classification to assess the damage from the targeted noise injection. Recall
that we try two attack variants, attack variant 1 and 2, where we inject one or
two mirrored sets of vertices and edges, respectively. This is to both i) understand
how much noise is needed to yield successful evasion, and ii) determine the cost
incurred by adding noise.

Spectral clustering generates 267 DGA clusters from the four malware families
across 12 days. Before the attack, only 0.4\% clusters (1 out of 267) are predicted with
the wrong labels. In comparison, after the attacks, all clusters are predicted with the
wrong labels. Next, we will examine the predicted class probabilities change in the true
class label.

Figure~\ref{fig:prediction_confidence} uses the violin plots to show the distribution of
predicted class probabilities for the \emph{true} DGA families, before and after the attacks.
The circle is the median value, the thick line within the violin indicates interquartile range,
and the thin line depicts a 95\% confidence interval. The shape of the violin shows the distribution
of the probability values. For example, the first violin in Figure~\ref{fig:prediction_confidence}
has a median of 100\% predicted class probabilities, and all data points in the interquartile range
have 100\% probability value.
Specifically, before the attacks, 238 clusters are predicted with 100\% class
probability that they belong to the true class, and only 28 clusters have a
probability between 60\% and 100\%. For example, the Pykspa cluster had a
class probability of only 10\% because it contained only two domain names that
had very different feature distributions from the majority of Pykspa clusters.
The two variants of the attack introduced \emph{at least} 50\% and 66\% noise
to the DGA clusters.

\paragraph{Minimal Knowledge} 
\label{par:Minimal Knowledge}

After the attacks, we classify each new adversarial cluster containing target
domains and plot the target class probability distributions in the
Figure~\ref{fig:prediction_confidence}. Attack variant 1 (``Minimal Benign DGA
1'') generated new clusters with $\leq 80\%$ predicted class probability, with
a median of 0\%. The predicted class probabilities of 84\% of the new clusters
drop to zero. Attack variant 2 further decreases the classifier prediction
confidence, as shown by ``Minimal Benign DGA 2'' in
Figure~\ref{fig:prediction_confidence}. After injecting two benign DGA domains,
the predicted class probabilities of 87\% of the new clusters plummet to 0\%.
The overall distribution of prediction confidences also shifts downward
compared to ``Minimal Benign DGA 1''.


\paragraph{Perfect Knowledge} 
\label{par:Perfect Knowledge}

The median of predicted class probabilities for DGA malware families drops to
10\%. As depicted by ``Perfect Long Tail 1'' in
Figure~\ref{fig:prediction_confidence}, 86\% of adversarial clusters were
assigned the probabilities of belonging to the true DGA class that are at most 10\% . The
distribution of class probability values has a smaller variance compared to
those in the ``Minimal Benign DGA 1''. ``Perfect Long Tail 2'' in
Figure~\ref{fig:prediction_confidence} shows that the maximum prediction
confidence is 30\%, slightly lower than the maximum 40\% confidence from the
targeted noise injection attack of ``Minimal Benign DGA 2''.


\paragraph{Moderate Knowledge} 
\label{par:Moderate Knowledge}

We see similar results for the two targeted noise injection attack variants in
the moderate knowledge case as in the other cases: a strong drop in predicted
class probabilities, with a smaller, more compact distribution of values for
attack variant 2. After attack variant 1, 98.3\% of new clusters were assigned
less than 20\% confidence; after attack variant 2, 98.8\% of new clusters have
less than 20\% confidence.


\emph{Spectral clustering can be largely defeated at all knowledge levels using
the targeted noise injection attacks.}

Since previous experiments show that minimal knowledge attackers can carry out
targeted noise injection as effectively as more powerful attackers, we will
simply demonstrate that the same targeted noise injection attack variant 1 in
minimal knowledge also works with community discovery and node2vec.

\subsubsection{Community Discovery}

We use the same set of DGA domains labeled in Spectral Clustering for evaluation.
Before the attack, 80\% clusters can be predicted with the correct label,
which dropped to 2\% after the attack.
Figure~\ref{fig:prediction_confidence_comm_n2v} shows the predicted class
probabilities for communities containing all \emph{target domains} before and
after the attack. Before the attack, the median of predicted probabilities is
90\%, and the interquartile range is from 50\% to 100\%.
Specifically, 71 communities contain target domains, among which ten
communities only contain one target domain, and seven communities have between
40\% to 70\% target domains. These noisy communities formed the lower part of
the distribution, with $\leq 50$\% predicted class probabilities in ``Community
Before Attack'', as shown in Figure~\ref{fig:prediction_confidence_comm_n2v}. After the
attack, the median class probability craters to 0\%. Overall 98\% of new
communities were predicted with lower than 50\% probability of belonging to the
true class, and 86\% of communities have lower than 10\% class probabilities.

{\it This demonstrates that the targeted noise injection attack is also
effective against the community discovery algorithm.}

\subsubsection{node2vec}

Using the same set of DGA domains labeled in Spectral Clustering,
before the attack, 89\% clusters can be predicted with the correct label,
which dropped to 0.8\% after the attack.
Figure~\ref{fig:prediction_confidence_comm_n2v} shows that, before the attack on
node2vec, the median of predicted probabilities is
100\%, and the interquartile range is from 90\% to 100\%.
A total of 85\% of clusters were predicted with at least 70\% class probability.
After the attack, 92\% clusters have at most 10\% predicted class
probabilities.

{\it Targeted noise injection attack also evades node2vec embeddings.}

\subsubsection{Targeted Noise Injection Costs} 
\label{ssub:Noise Injection Costs}

\begin{table}
  \centering
  \begin{tabular}{lrr}
    \textbf{Before Attack} & \multicolumn{2}{ c }{\textbf{< 95th Percentile, 9.12\% of hosts}} \\
    Average Increase & From Percentile & To Percentile \\
    \hline
     Attack Variant 1 & 69.86\% & 88.73\% \\
     Attack Variant 2 & 69.86\% & 93.98\% \\
    \hline
    \textbf{Before Attack} & \multicolumn{2}{ c }{\textbf{>= 95th Percentile, 90.88\% of hosts}} \\
     Average Increase & From Percentile & To Percentile \\
    \hline
     Attack Variant 1 & 99.74\% & 99.85\%  \\
     Attack Variant 2 & 99.74\% & 99.88\% \\
  \end{tabular}
  \caption{Anomaly cost as percentile of the distinct number of NXDOMAINs queried
  by hosts, before and after the attack. Only 9.12\% of infected hosts
  become more suspicious, while the rest remain the same.}
  \label{tab:anomalycost}
\end{table}

It is simple for malware to query additional domains, however, infected hosts
engaging in such queries may become more suspicious and easier to detect due to
the extra network signal they produce. This may cause the anomaly cost of the
targeted noise injection attack to be high enough to render it useless.

We analyze the anomaly cost by measuring the infected host percentile of the
NXDOMAIN distribution both before and after the attacks for the two
variants of the targeted noise injection attacks, summarized in
Table~\ref{tab:anomalycost}. Before any attack, only 9.12\% of infected hosts
ranked lower than $95^{th}$ percentile, and the remaining 90.88\% of them ranked
higher than $95^{th}$ percentile. This means that, without any attack, infected
hosts were already querying more unique NXDOMAINs than most hosts in the
network. However, doing targeted noise injection attacks further increases the
percentile ranks of the infected hosts, but not substantially.

We separated the results based on whether infected hosts were querying fewer
domains than 95\% of all hosts in the local network.
Table~\ref{tab:anomalycost} shows that among the 9.12\% infected hosts ranked
lower than $95^{th}$ percentile before the attack, they increased from an
average percentile of 69.86\% to 88.73\% after the targeted noise injection
attack variant 1. Furthermore, they increased to 93.98\% after attack variant
2. However, 90.88\% of infected hosts did not become more anomalous. They
were ranked higher than the $95^{th}$ percentile before the attack. Their
average percentile increased by 0.11\% after attack variant 1, and by 0.14\%
after attack variant 2. Because they were querying more domains than other
hosts before the attack, injecting noise does not change their percentile
substantially.

{\it The majority of hosts had little change in ``suspiciousness'', whereas a
small percentage of hosts increased their suspiciousness after the targeted
noise injection attacks.}



\subsection{Small Community} 
\label{sub:Diffusion}

We choose a group of 618 domains and 10 infected hosts belonging to Suppobox as
the basis for the small community attack. They form a community using the
community discovery algorithm, and two clusters using spectral or node2vec
embeddings. A small community attack is successful if and only if \emph{all}
DGA domains join either the ``death star,'' or clusters where the
subsequent classifier does not predict them as the true malware DGA class.
Recall that the death star cluster contains tens of thousands of domains that cannot
be properly classified. In all experiments, the small community plots denote
the configurations where an attack succeeds based on the aforementioned
criteria. This is represented by green regions (see
Figure~\ref{fig:enum_multiplot}) when the ``death star'' is joined, or white
cells when the noisy clusters cannot be predicted as the true class label (see
Figure~\ref{fig:n2v_haslabel_multiplot_contextsize}) when using node2vec.

\subsubsection{Spectral Clustering}

As described earlier, the small community attack can only be verified in the
perfect and moderate knowledge cases. In the minimal knowledge case, however,
an attacker can still mount the attack by randomly removing edges and nodes, as
described in Section~\ref{fig:diffusion_example}, while hoping for the best.

\paragraph{Minimal Knowledge} 
\label{par:Minimal Knowledge}

\begin{figure}[t]
    \centering
    \scalebox{0.48}{\includegraphics{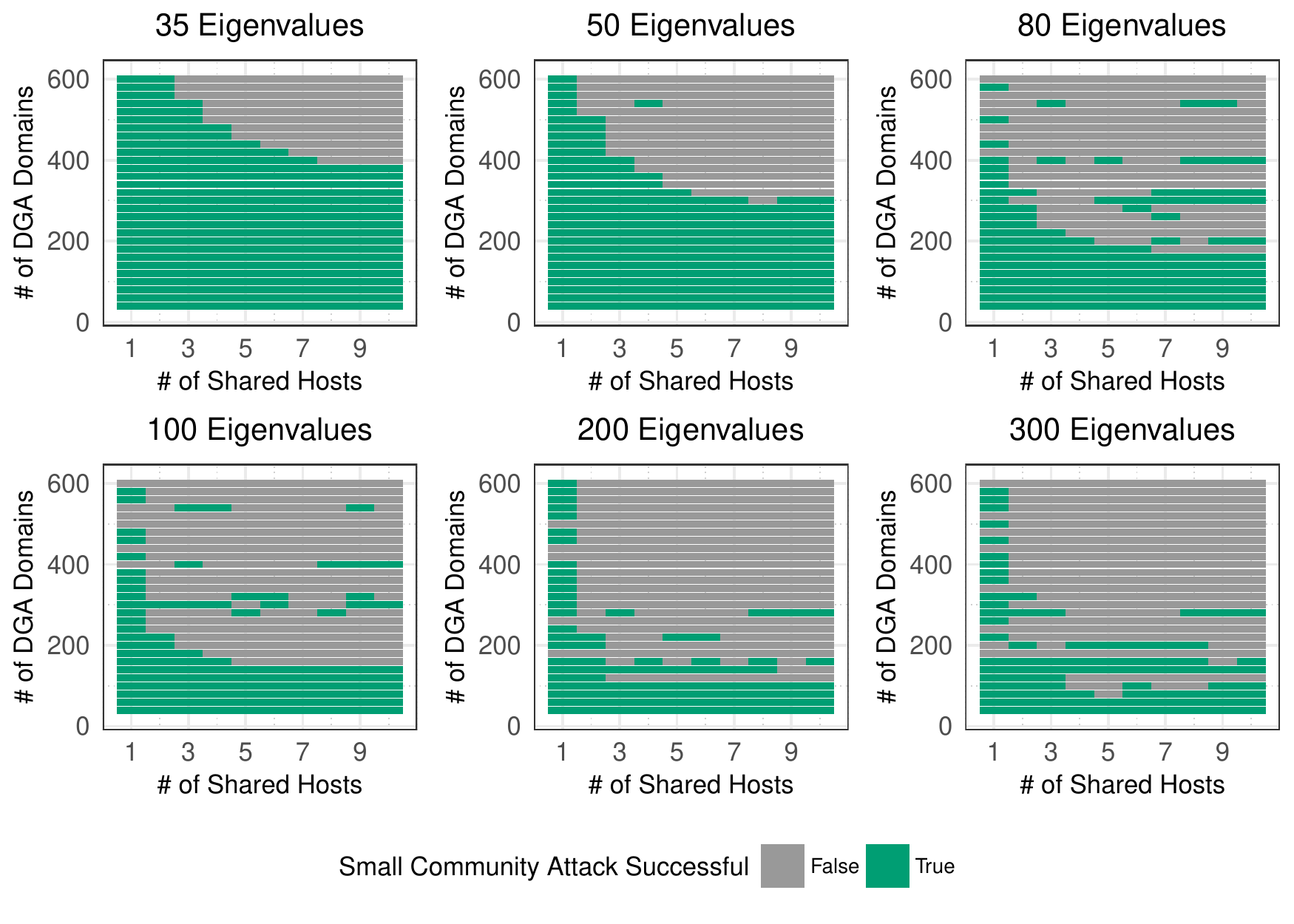}}
    \caption{Different number of eigenvalues.}
    \label{fig:enum_multiplot}
\end{figure}

The upper-leftmost plot in Figure~\ref{fig:enum_multiplot} demonstrates the
possible successful configurations for mounting the small community attack by
randomly removing nodes and edges. The plot shows the remaining number of
NXDOMAINs on the Y-axis ($|V| - n_v$) and the remaining number of connections
from infected hosts for each NXDOMAIN on the X-axis ($|U| - n_e$).  The shaded
region shows approximately a 75.16\% success rate for an attacker with no
knowledge of the defender's graph $\mathcal{G}$. While a minimal knowledge
attacker cannot guarantee their attack will succeed, they nonetheless have a
high chance of success.


\paragraph{Perfect Knowledge} 
\label{par:Perfect Knowledge}

The upper-left plot in Figure~\ref{fig:enum_multiplot} depicts a successful
small community attack area when the computed SVD rank is 35.  The figure shows
that only a small set of configurations with 380 to 618 DGA domains, each
queried by between 3 to 10 random hosts, were unable to successfully launch a
small community attack. The cost of the small community attack is very low
against the system, which is configured with rank 35 and runs in this network. For
example, an adversary controlling the DGA does not need to give up any
infection, but only needs to reduce the number of infected hosts that query a
common set of DGA domains from 618 to 380 in order to hide
the domains. In this case, by removing $n_v = 238$ NXDOMAINs, the attacker does
not lose any additional host querying activities $min(n_e) = 0$.  But if the
attacker needs extra redundancy provided by 460 distinct NXDOMAINs, each domain
can only be queried by a subset of 5 hosts. Then $n_v= 158$, and accordingly, $min(n_e)
= 5$. In this case, the attacker does not need to lose control of any infected
hosts, but she does need to coordinate each five infected hosts to query a subset
of distinct NXDOMAINs that do not overlap with each other.


\paragraph{Moderate Knowledge}

\begin{figure}[t!]
    \centering
    \scalebox{0.3}{\includegraphics{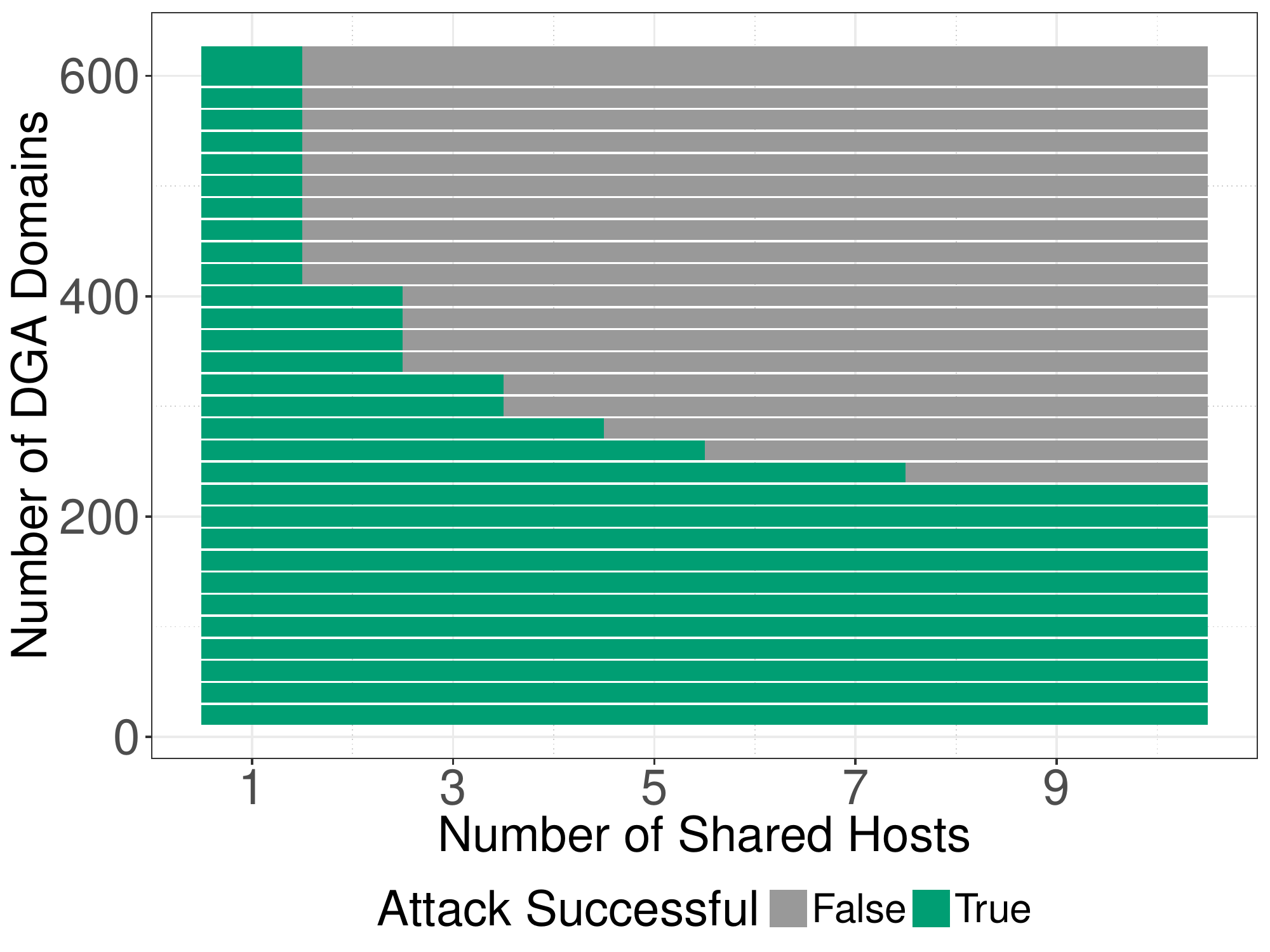}}
    \caption{Success area for joining the death star of the surrogate dataset in the
    moderate knowledge case. All the successful attack configurations worked in the
    ground truth network.}
   \label{fig:surrogate}
\end{figure}

After reducing the number of DGA domains and the number of infected hosts per
domain to the successful attack area shown in Figure~\ref{fig:surrogate}, the
DGA domains join the surrogate death star. We test that these values also work
to join the original death star. Because the original network size is larger
than the surrogate network size, the real successful area (the top-left plot in
Figure~\ref{fig:enum_multiplot}) is much bigger than the one shown
Figure~\ref{fig:surrogate}. Thus, the small community attack works with
moderate knowledge when the surrogate dataset is a smaller network than the
original network. If the surrogate dataset is a larger network, the adversary
may miscalculate the cost of joining the death star, which may not work in the
original network. By using such a surrogate dataset, the adversary will likely
choose fewer DGA domains and their shared hosts to simulate a successful
attack, compared to the ideal case in perfect knowledge. In other words, the
practical cost of launching a small community attack with moderate knowledge is more
than the minimal cost of such an attack in the original network. We explore the effect of
network size in Section~\ref{section:size_of_network}.

\emph{Spectral clustering can be evaded using the small community attack, even
when the attack cannot be verified by the attacker with a success rate of
75\%+. More sophisticated attackers can always evade.}

\subsubsection{Community Discovery}

Unlike graph embedding techniques that lose information about smaller
components of the graph, community discovery algorithms do not lose information
and can properly handle portions of $\mathcal{G}$ with exactly one edge. Rather
than clustering poorly with other small components, they are considered to be
separate communities. So the cost of the small community attack is much higher
than with graph embeddings because attackers must generate singletons that are
small enough to evade classification, forcing the attacker's graph to be
disconnected. Therefore, they can evade clustering with the cost of losing their ability to
efficiently manage their bots. For example, to evade community discovery in the example
presented in Figure~\ref{fig:diffusion_example}, an attacker would have to use
the modified attack graph $G'_{4}$ and the drop from $D(G)=0.5$ to
$D(G'_{4})=0.25$ is enough to consider the attack cost too high. In the DGA
case, this would mean each infection would need its own distinct domain-name
generation algorithm, which would be an exceedingly high cost for an attacker. As such, we do
not compute results for small community attacks on community discovery.

\emph{Community discovery is resistant to the small community attack due to the
high costs it would cause the attacker, however, spectral methods and node2vec
are more likely to be used by defenders as they result in cleaner clusters and
better classification results.}

\subsubsection{node2vec}

\begin{figure}[t]
    \centering
    \scalebox{0.48}{\includegraphics{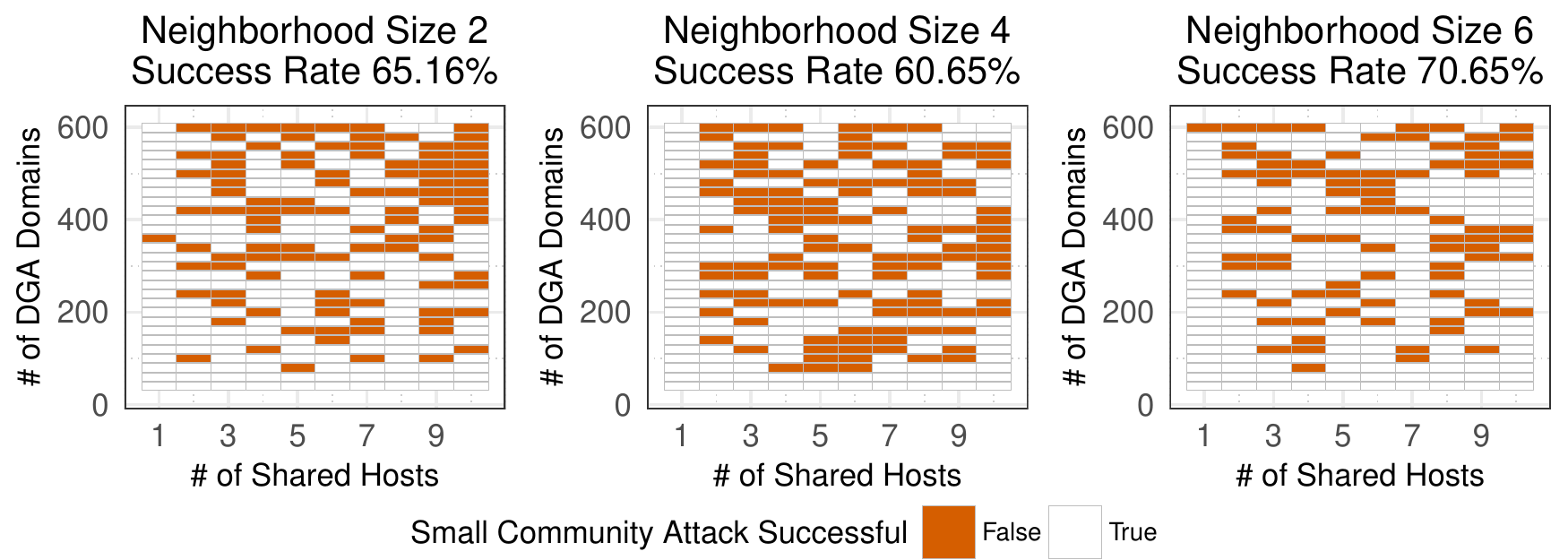}}
    \caption{Success area of small community attacks with different context size.}
    \label{fig:n2v_haslabel_multiplot_contextsize}
\end{figure}

The third plot in Figure~\ref{fig:n2v_haslabel_multiplot_contextsize} shows
that the small community attack is still possible with node2vec, using
aforementioned hyperparameters (Section~\ref{sub:parameters}). The attack is
possible when the number of shared hosts is 1 (the first column except the top
cell), and when the number of DGA domains is $\leq 40$ (the bottom two rows).
Elsewhere, the attack succeeds randomly due to the random walk.  In summary,
the small community attack is definitely possible with very small component
sizes. Compared to SVD, the cost is higher here. For example, the attacker
needs to give up $n_v = 578$ unique NXDOMAINs in a day, along with $n_e=0$, for
the small community attack to be successful. But if the attacker is not willing
to give up such cost, the small community attack is not guaranteed to succeed
given the randomness of neighborhood sampling. However, if a minimal knowledge
attacker randomly chooses any $n_v$ and $n_e$ for a small community attack, she
will have a 70.65\% attack success rate shown by the third plot in
Figure~\ref{fig:n2v_haslabel_multiplot_contextsize}.

\emph{node2vec is susceptible to the small community attack, but with fewer
guarantees and higher costs than in the spectral case, due to its inherent
randomness. node2vec being used in Pleiades would render the system more
resilient against small community attacks.}

\subsubsection{Small Community Costs} 
\label{ssub:Diffusion Costs}

The cost of the small community attack is affected by both the size of network
and change in density when the attack is performed.

\begin{figure}[t!]
    \centering
    \scalebox{0.48}{\includegraphics{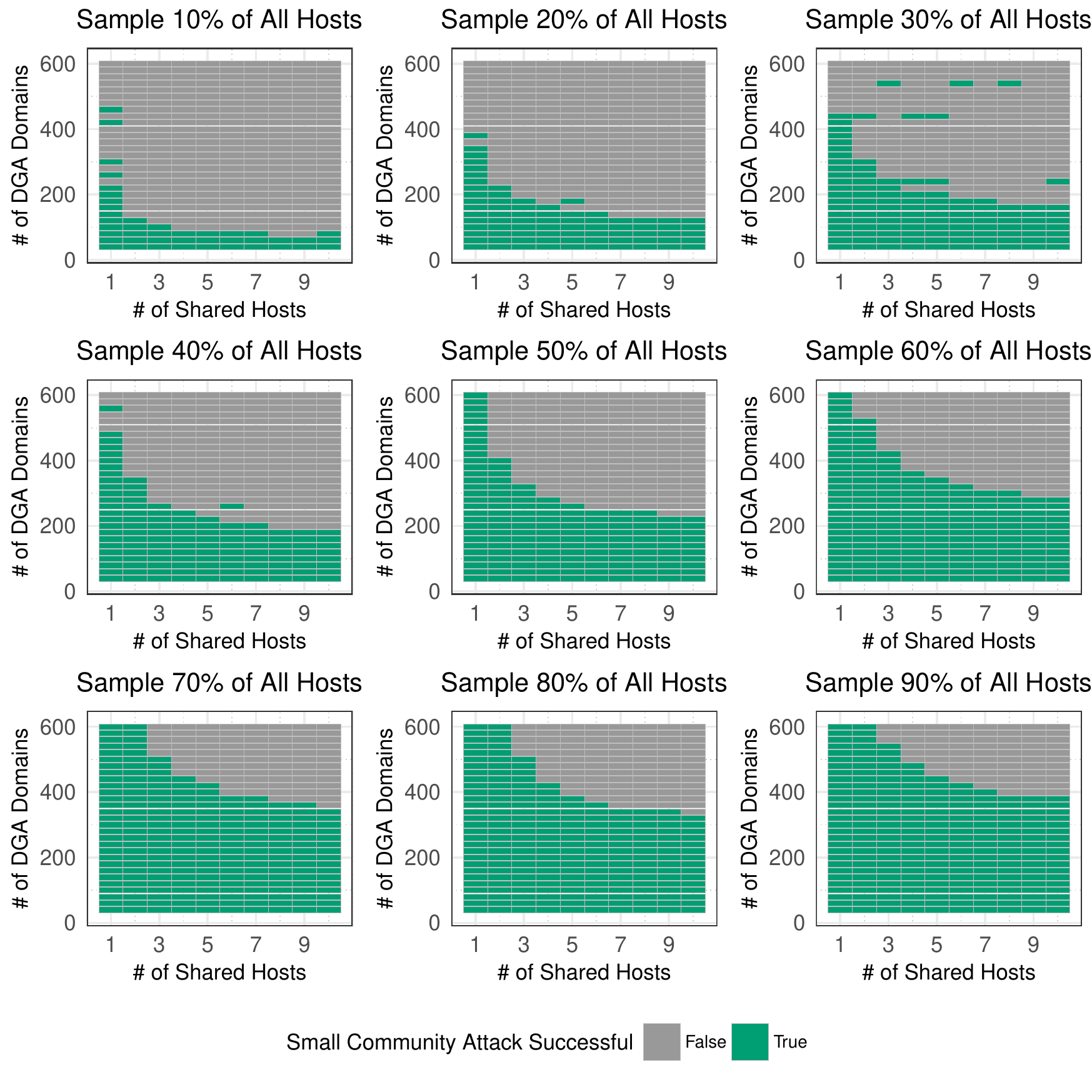}}
    \caption{
    Different sizes of the network dataset.
    }
     \label{fig:networksize}
\end{figure}

\paragraph{\textbf{Size of Network}}
\label{section:size_of_network}

The network size is related to the number of nodes (hosts and domains) and the
number of edges (the query relationship). As a straightforward way to model the
network size, we randomly sample the hosts in the ground truth network dataset
along with all domains queried. We also keep the same attacker subgraph $G$,
containing the Suppobox DGA community with 10 infected hosts and 618 DGA
domains, along with other domains queried by these hosts for the experiment.

Figure~\ref{fig:networksize} shows the small community attack results by
sampling 10\% to 90\% of all hosts. When only 10\% of hosts were sampled, the
small community attack failed in most areas of the plot. The attack success
area increases as the network size gets larger. This means that the cost for
small community attack is lower in a larger network than in a smaller network,
given the same hyperparameters. A larger network is harder to accurately
represent in an embedding, which provides more areas for attackers to hide and
evade.

\emph{A moderate knowledge level attacker should attempt to acquire a surrogate
network dataset smaller than the ground truth network dataset for a safe
estimate of their small community attack cost.}

\paragraph{\textbf{Agility Cost}}

\begin{table}
  \centering
  \begin{tabular}{lrr|r}
        \hline
      & \multicolumn{3}{ c} {\textbf{Spectral Clustering}} \\
      & \multicolumn{2}{ c}{Density}  & \\
     Join Death Star & Median & Maxium & Minimum Cost \\
    \hline
     SVD rank 35 & 0.078  & 0.61 & 0 \\
     SVD rank 50 & 0.11 & 0.45 & 0.03 \\
     SVD rank 80 & 0.065 & 0.26 & 0.22  \\
     SVD rank 100 & 0.052  & 0.19 & 0.29 \\
     SVD rank 200 & 0.0032 & 0.10 & 0.38 \\
     SVD rank 300 & 0.026 & 0.26 & 0.22 \\
     \hline
     \hline
     & \multicolumn{3}{ c} {\textbf{node2vec}} \\
      & \multicolumn{2}{ c}{Density}  & \\
     & Median & Maxium & Minimum Cost \\
     \hline
     Neighborhood Size 6 & 0.026 & 0.065 & 0.415 \\
     \hline
  \end{tabular}
  \caption{Agility cost of small community attacks under different
  hyperparameter configurations.}
  \label{tab:agilitycost}
\end{table}

By removing nodes and edges, the attacker loses redundancy. For example,
hosts need to query fewer DGA domains, or malware can be allowed fewer
malicious actions. We measure the agility cost by the change in density of the
attack graph. Density captures the number of edges present in the attack graph
over the maximal number of possible edges. In Section~\ref{sub:Diffusion},
Equation~\ref{eqn:density} defines the attack graph density before the small
community attack; and Equation~\ref{eqn:attackdensity} defines the density
after the attack. Before the attack, $D(G) = 0.48$ for the Suppobox
community. For each SVD rank parameter, we record attack configurations that
were successful small community attacks as green ares in
Figure~\ref{fig:enum_multiplot}.  There are some outliers outside the
continuous area. Although these attacks do not make NXDOMAINs join the death
star, they move NXDOMAINs to clusters that cannot be predicted with the correct
label. To measure the minimum agility cost, we exclude the outliers by only
calculating attacker graph density that resulted in joining the death star.
Table~\ref{tab:agilitycost} summarizes the median and maximum attacker graph
density in these small community attacks, with the minimum cost represented by the difference
between $D(G)$ and $max(D(G'))$. When the SVD rank is 35, the $max(D(G'))$ to
join the death star is slightly bigger than $D(G)$, which means there is no
cost in launching the small community attack. In this case, the attacker can
evade while having \emph{more} connectivity. As the SVD rank increases, the
attacker graph density is reduced, which means a successful attack is more
costly to the adversary. Also, the minimum cost increases as the SVD rank
increases. For example, when the SVD rank is 80, the minimum cost is
0.22, reducing the attack graph density from 0.48 to 0.26. The attacker needs
to reduce the number of distinct DGA domains from 618 to 160 to evade, but each
domain can be queried by all infected hosts.  In comparison, when the SVD rank
is 200, the minimum cost is 0.38.  The attacker needs to further reduce the
number of distinct DGA domains to 60 to evade, with each domain queried by all
infected hosts.  The attack graph density is reduced from 0.48 to 0.1, losing
79\% ($\frac{0.38}{0.48}$) of queries to distinct DGA domains. This means that
tuning hyperparameters can increase the small community attack cost and
potentially render this attack ineffective.

Similarly, for node2vec, the minimum cost of a certain small community attack
is higher than spectral clustering. We compute the attacker graph density only
for the white area in Figure~\ref{fig:n2v_haslabel_multiplot_contextsize}
without randomness, i.e., the first column and bottom two rows. In contrast to
spectral clustering, node2vec requires a much higher minimum cost for a
guaranteed small community attack, which indicates that node2vec is more
resilient to this attack.
The smallest communities in Figure~\ref{fig:n2v_haslabel_multiplot_contextsize}
(i.e., the first column and bottom two rows)
are likely undersampled, because choosing 15 walks per node and walk length 20
using cluster validity in Section~\ref{section:node2vec_validity} prefers labeled DGA communities
that are relatively bigger, which makes few neighborhood observations for
extremely small islands insignificant, and thus allows small community attacks.
Note the randomness in the remaining portion of the
plot.
Since node2vec uses the random walk process to sample the neighborhoods
of all nodes, there exists randomness in the neighborhood observations.
This shows that the randomness inherent to node2vec makes the attacks
succeed at random in the remaining portion of Figure~\ref{fig:n2v_haslabel_multiplot_contextsize}.
This both suggests a system like Pleiades would benefit from
node2vec to reduce the guarantee of attacks, as well as allow a defender to
identify if an attacker is evading by chance encounters where the evasion fails
over time. While the minimum attack cost is the same with different
neighborhood sizes for a guaranteed successful attack, the attack success rate changes.
The neighborhood sizes 2, 4, and 6 have attack success rate 65.16\%, 60.65\%, and 70.65\%
respectively (Figure~\ref{fig:n2v_haslabel_multiplot_contextsize}).
We will discuss how we can use different hyperparameters to further reduce the
success rate of the small community attack in Section~\ref{sub:Improving
Hyperparameter Selection}.

\emph{These costs further demonstrate node2vec's superiority over spectral
clustering in resisting small community attacks.}




\section{Defense}

Since the noise injection attack and the small community attack violate the
fundamental assumptions used by graph clustering techniques, it is very hard to
completely eliminate the problem. In this section, we propose two defense
techniques that help Pleiades retain its detection capabilities against the two
attacks. The first one is to train the classifier with noise, which remediates
the noise injection attack to some extent. The second one is to use the small
community attack as an adversarial guideline to choose better hyperparameters for
graph embeddings, which increases the cost of launching a successful small
community attack.

\subsection{Training Classifier with Noise}

\begin{table}
  \centering
  \begin{tabular}{lrrrr}
     & \multicolumn{4}{ c }{False Positive Rate} \\
    \textbf{Model} & Pykspa & Gimemo & Suppobox & Murofet \\
    \hline
     Original & 0.32\% & 0.29\% & 0\% & 0\% \\
     Model A & 1.64\% & 0.39\% & 0.10\% & 0\% \\
     Model B & 1.62\% & 0.10\% & 1.23\% & 0.30\% \\
     Model C & 1.46\% & 1.17\% & 1.23\% & 0\% \\
    \hline
  \end{tabular}
  \caption{False Positive Rate for four DGA families before retraining, and after retraining
  with three types of noise.}
  \label{tab:fpr}
\end{table}

By retraining the classifier, it becomes more resistant to noise that could be
injected by the adversary in the unsupervised phase of Pleiades.  We used
domains from the benign DGA to poison the clusters of malicious DGAs. We
retrained the classifier using clusters generated by the noise injection attack
variant 1 (``Minimal Benign DGA 1'', $m=1$, Algorithm~\ref{alg:noise} in Section~\ref{sub:Noise Injection})
from SVD, yielding \emph{model A}. We
tested model A against the adversarial clusters generated by the same noise
injection attack under community discovery and node2vec. The first two violins
in Figure~\ref{fig:retrain} show that model A increases the overall predicted
class probabilities compared to the ``After Attack'' violins in
Figure~\ref{fig:prediction_confidence}. In community discovery, the
accuracy increased from 2\% to 98\%; and in node2vec, the accuracy increased
from 0.8\% to 98\%. To summarize, retraining with noisy clusters containing a
benign DGA from SVD can remediate the same attack on community discovery and
node2vec. We see this same effect even when the noise levels are
doubled ($m=2$, Algorithm~\ref{alg:noise} in Section~\ref{sub:Noise Injection}). When
models were trained with half the noise ($m=1$, Algorithm~\ref{alg:noise} in Section~\ref{sub:Noise Injection}),
they were able to more accurately predict the correct label.
Among them, only an average of 7.3\% clusters
are predicted with the wrong labels, decreased from 100\% before retraining.

In comparison with Figure~\ref{fig:prediction_confidence}, the average
prediction confidence increased significantly. Before retraining, the average
prediction confidence of ``Minimal Benign DGA 2'', ``Moderate 2'', and
``Perfect Long Tail 2'' are 10\%, 20\%, and 20\%. After retraining, they
increased to 70\%, 90\%, and 80\%, respectively. The accuracy of the models
remain roughly the same before and after retraining. However, retraining with
noisy clusters increased the false positive rate in most cases
(Table~\ref{tab:fpr}).

It is important to note that this defense only trains the classifier with noise
that has been witnessed. New noise will appear, but the fundamental attack on
the unsupervised component remains the same. Therefore, defenders will be
alerted by plumetting accuracies in their models.  Our defenses are simple and
future work should be done to make clustering systems more robust.

\subsection{Improving Hyperparameter Selection} 
\label{sub:Improving Hyperparameter Selection}

\begin{figure}[ht!]
	\centering
        \subfloat[]
        {\includegraphics[scale=0.4]{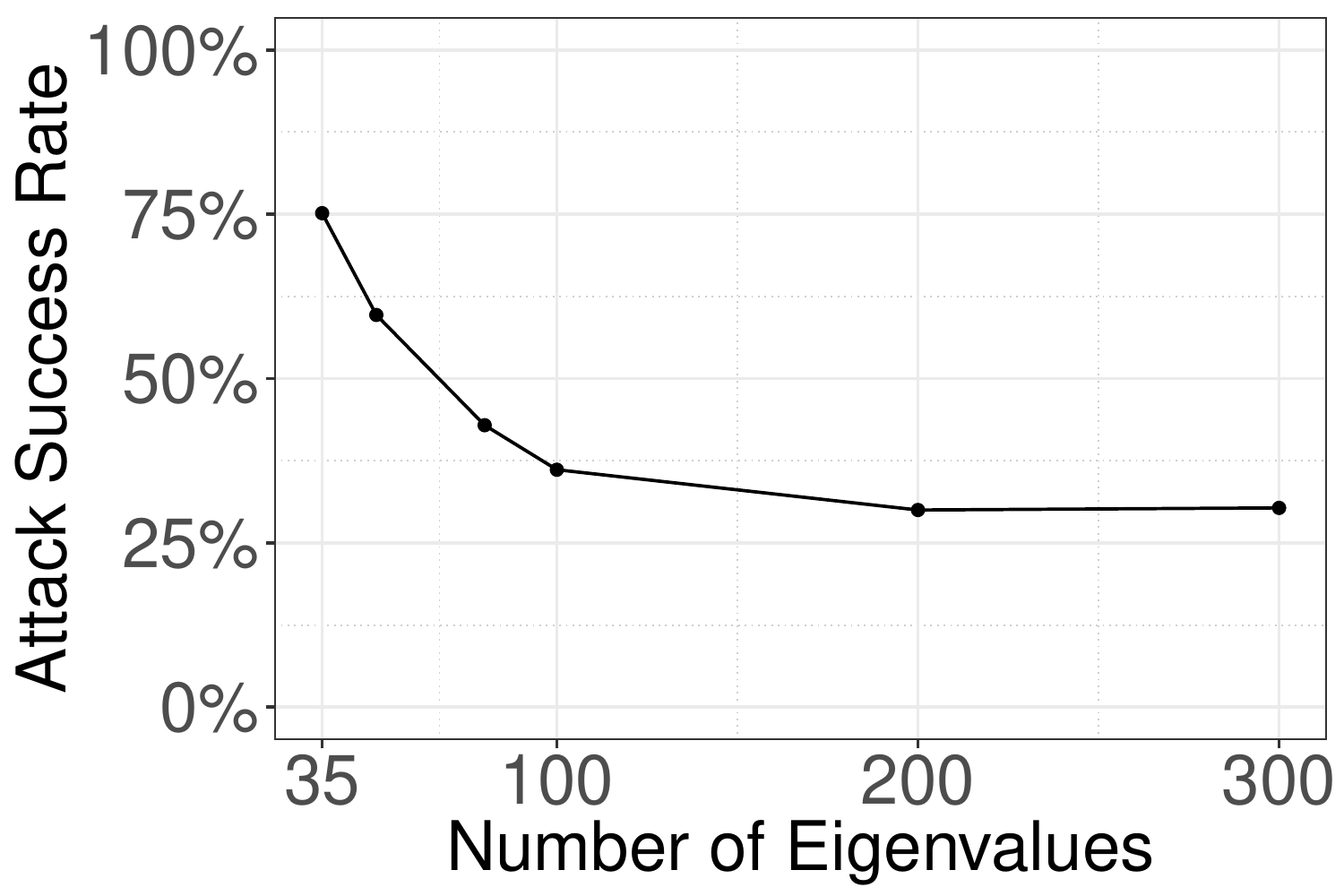}
        \label{fig:enumds}}

        \subfloat[]
        {\includegraphics[scale=0.4]{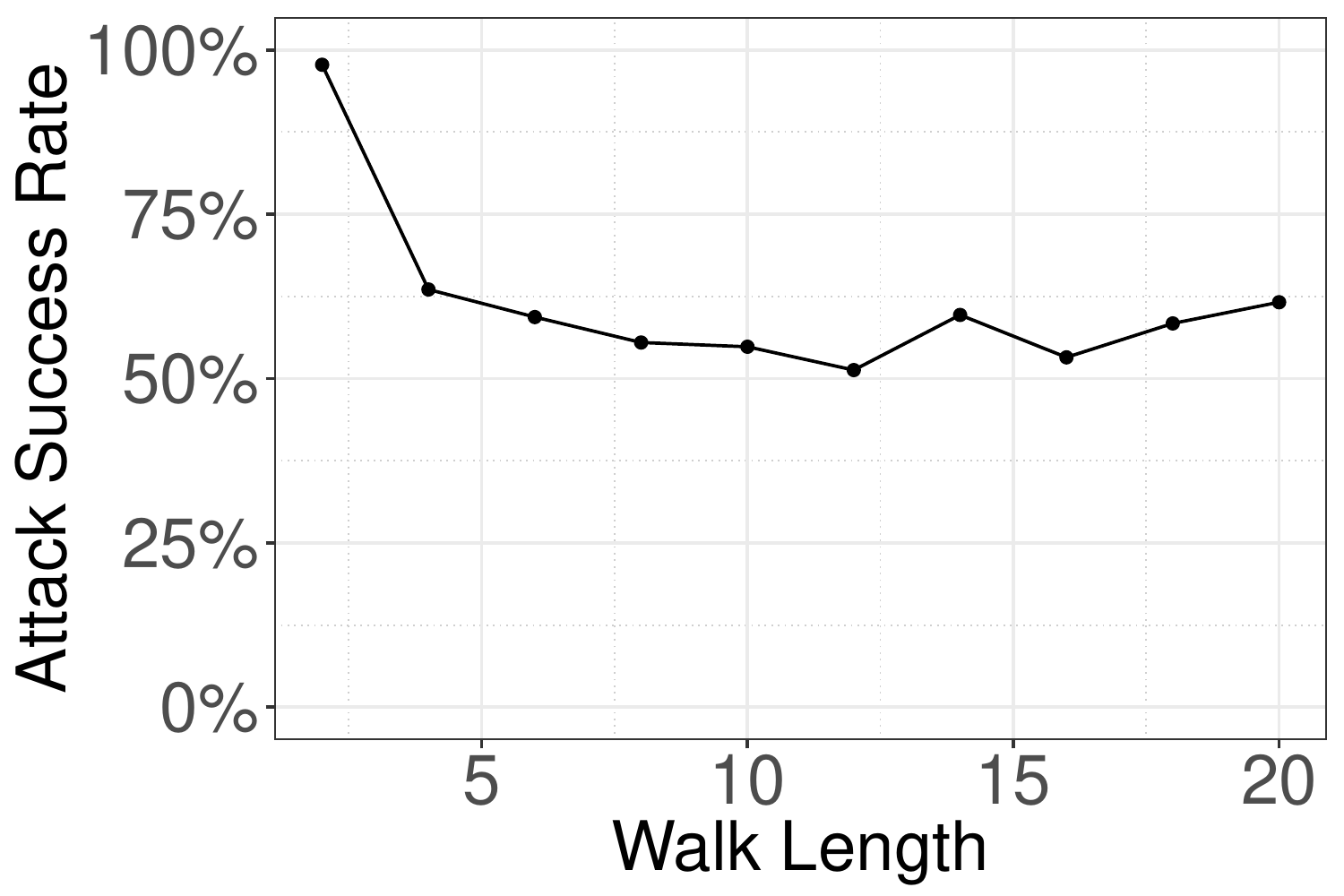}
        \label{fig:n2v_choose_walklength}}

        \caption{
	Figure~\ref{fig:enumds}:
        Using the small community attack to choose the number of eigenvalues for SVD.
        Figure~\ref{fig:n2v_choose_walklength}: Using the small community attack to choose the length of walk for
        node2vec.
        }
        \label{fig:parameters}
\end{figure}

Small community attacks show that the traditional ways of choosing
hyperparameters (Section~\ref{sub:parameters}) is not enough when facing
adversaries. Luckily, the small community attack can be used to choose more
resistant hyperparameters. We show that better selection can reduce the number
of successful small community attack instances from our previous experiments.

We plot the successful attack rate under different number of eigenvalues in
Figure~\ref{fig:enumds}. The successful attack rate decreases as the number of
eigenvalues computed increases, and the line plateaus after 200 eigenvalues. It
means that a defender running Pleiades should select the first 200 eigenvalues,
instead of 35 indicated by the scree plot in Figure~\ref{fig:screeplot}. If we
use the small community attack in this way, we can choose better parameters for
the system and also know under which parameters the system is vulnerable.

Similarly, for node2vec, using the small community attack to choose
hyperparameters can reduce the attack success rate. The cluster validity
metrics suggest we choose neighborhood size six, and walk length of 20.
However, if we evaluate the graph clustering using the success rate of the
small community attack, these hyperparameters are not optimal. First, for the
neighborhood size, Figure~\ref{fig:n2v_haslabel_multiplot_contextsize} shows
that a smaller neighborhood size of four introduces a lower attack success
rate. Second, we plot the attack success rate under different walk lengths in
Figure~\ref{fig:n2v_choose_walklength}. This figure shows that a walk length of
12 is preferred over 20, because the former allows 51.29\% attack success rate
compared to 61.61\% of the latter. In other words, the smaller neighborhood
size and shorter walk length can tolerate the small community attack better,
presumably because they do not oversample larger communities with more distinct
neighborhood observations. In other words, smaller communities are not
undersampled.
We recommend using the small community attack success rate to evaluate
the clustering hyperparameter selection, in addition to traditional cluster validity
indices.


\section{Discussion}

We acknowledge that details surrounding the implementation of the attacks are
specific to Pleiades, however, the graph representation suggests the attacks
may work on other graph-based systems. In this section, we briefly discuss
issues to consider to generalize the attacks.

Nodes and edges can be trivially injected or removed in the graph Pleiades uses,
which are generated by malware resolving domain names. In other security contexts, the
set of injectable/removable nodes varies. It is possible that some nodes and
edges must exist in order for certain attack actions to succeed. For example, a
phishing email using a malicious attachment requires at least the \emph{read}
system call to successfully infect a host, which cannot be removed from the
system call graph. On the other hand, it can be difficult to add certain nodes
and edges. Therefore, in addition to the anomaly cost (Section~\ref{ssub:Noise
Injection Costs}) and agility cost (Section~\ref{ssub:Diffusion Costs}), the
action of graph manipulation itself has costs depending on the data that
underlies the graph representation. This should be carefully considered when
generalizing the attacks to other systems and we leave this to be future work.
Tighter costs may exist, but our approaches point in a promising direction.

\section{Conclusions}

We have demonstrated that generic attacks on graphs can break a real-world
system that uses several popular graph-based modeling techniques. These attacks
can often be performed by limited adversaries at low cost; however, simple
defenses can reduce their effectiveness or likelihood of success. To summarize
how defenders can improve their systems: hyperparameter selection should be
optimized for reducing the success rate of small community attacks, and
retraining can be used to lessen the impact of noise injection attacks.
Furthermore, state of the art graph embedding techniques like node2vec appear
to be more resistant against small community attacks, which suggests Pleiades
and other systems would be harder to adversarially manipulate using node2vec
over community finding, or spectral methods (see
Figure~\ref{fig:n2v_haslabel_multiplot_contextsize} vs.
Figure~\ref{fig:enum_multiplot}).

\section{ACKNOWLEDGMENTS}

We thank our anonymous reviewers for their invaluable feedback, and
Dr. Rosa Romero-G\'{o}mez for her help in visualization.
This material is based upon work supported in part by the US
Department of Commerce grants no. 2106DEK and 2106DZD, National
Science Foundation (NSF) grant no. 2106DGX
and  Air Force Research Laboratory/Defense Advanced Research
Projects Agency grant no. 2106DTX. Any opinions, findings,
and conclusions or recommendations expressed in this
material are those of the authors and do not necessarily reflect the
views of the US Department of Commerce, National Science Foundation,
Air Force Research Laboratory, or Defense Advanced Research
Projects Agency.

\balance
\bibliographystyle{ACM-Reference-Format}
{
\bibliography{advdga}
}
\clearpage
\section{Appendices}

\subsection{Unique Domains queried by Hosts}
\label{sub:cdf}

\begin{figure}[t]
    \centering
    \scalebox{0.3}{\includegraphics{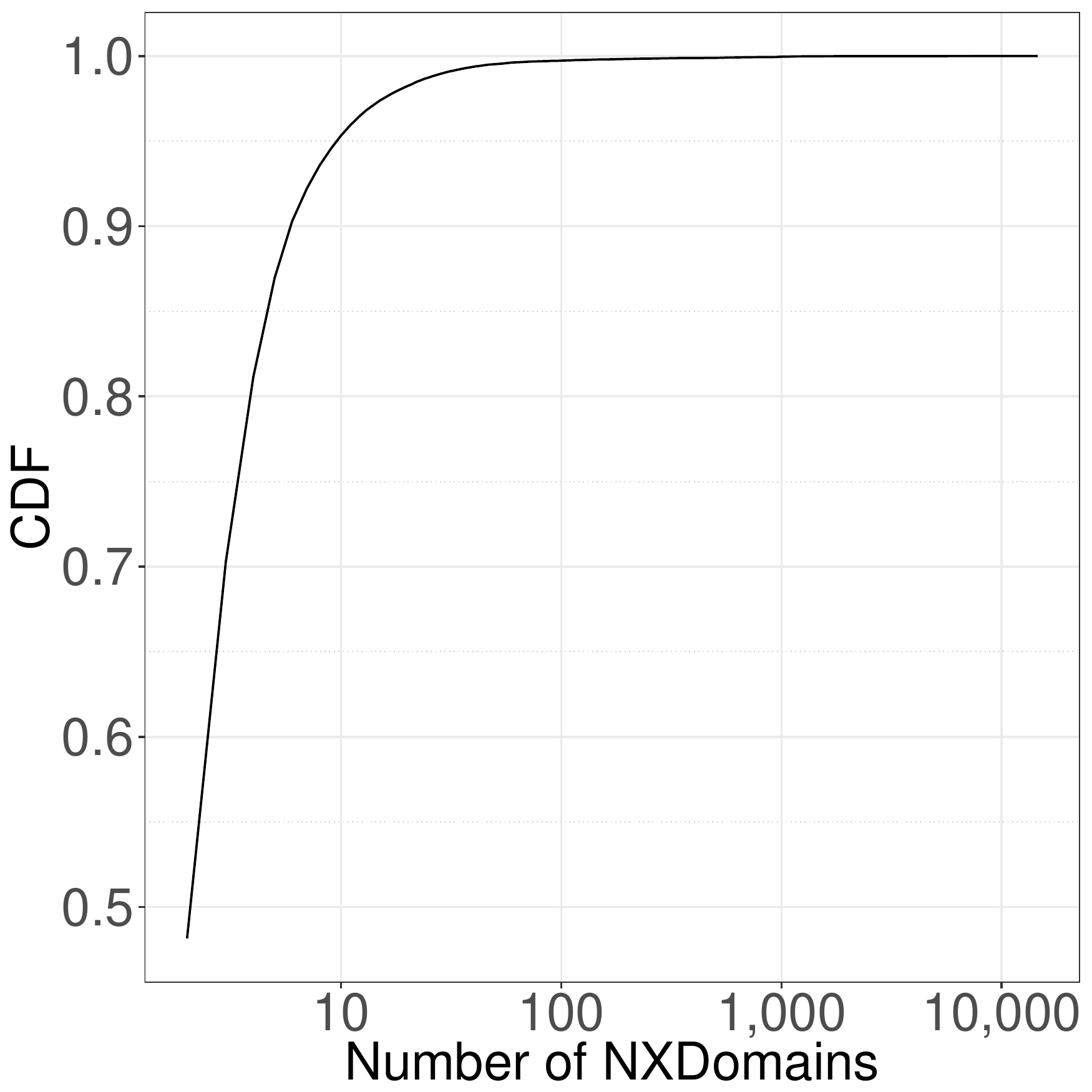}}
    \caption{
    Cumulative distribution of distinct number of NXDOMAINs queried by
    each host in 12/18/2016.
    }
   \label{fig:nxnum_per_host_cdf}
\end{figure}

Figure~\ref{fig:nxnum_per_host_cdf} shows the
cumulative distribution for distinct number of NXDOMAINs queried by hosts seen
on 12/18/2016 in the network datasets from the telecommunication network. The
CDF shows that a host querying two distinct NXDOMAINs is at the $48^{th}$
percentile, and a host querying 10 distinct NXDOMAINs is at the $95^{th}$
percentile. 

\subsection{Labeled DGA Families} 
\label{sub:dgalabels}

We use default parameters to generate different versions of the malware
families for 18 different seed dates. The number of domains generated
for each malware family is recorded in the top part of
Table~\ref{tab:dga-families}.

\begin{table}
  \centering
  \begin{tabular}{lrr}
    \textbf{DGA Family} & \textbf{\# of Domains} & \textbf{\# of Feature Vectors} \\
    \hline
    \hline
    Chinad & 4,608 & 18 \\
    Corebot & 720 & 18 \\
    Gozi & 864 & 72 \\
    Locky & 360 & 36 \\
    Murofet & 54,720 & 56 \\
    Necurs & 36,864 & 18 \\
    NewGOZ & 18,000 & 18 \\
    PadCrypt & 1,728  & 36 \\
    Qadars & 3,600 & 18 \\
    Qakbot & 180,000 & 35 \\
    Ranbyus & 720 & 18 \\
    Sisron & 739 & 19 \\
    Symmi & 1,152 & 18 \\
    Pykspa & 90,300 & 48 \\
    \hline
    Pykspa & 1,190 & 40 \\
    Gimemo & 9,144 & 17 \\
    Suppobox & 12,846 & 40 \\
    \hline
  \end{tabular}
  \caption{DGA families contained within our ground truth dataset.}
  \label{tab:dga-families}
\end{table}


\subsection{Reimplementing Pleiades}
\label{section:pleiades}

We implement a simplified version of the Pleiades DGA detection
system. We follow the exact next steps to implement the graph clustering and
modeling components of Pleiades. 

\begin{enumerate}

  \item From the NXDOMAIN query data, we filter out hosts that only queried one
	  domain name in a day (as the authors of Pleiades did). 

  \item We construct an association matrix representing the bipartite graph
    between hosts and the NXDOMAINs they queried.  Each row represents one host
    and each column represents one NXDOMAIN.  If host $i$ queried NXDOMAIN $j$
    in that day, we assign weight $w_{ij} = 1$ in the matrix. Otherwise, we
    assign $w_{ij} = 0$. Then, each row is normalized such that the sum of
    weights is one.

  \item Next, we do Singular Value Decomposition (SVD) over this matrix and
    keep the first $N$ eigenvalues.  For our dataset, we choose $N=35$
    according to the scree plot of Eigenvalues.  Figure~\ref{fig:screeplot}
    shows that the Eigenvalues line plateaus after $N>=35$.

  \item The resulting eigenvectors are used for XMeans clustering.

  \item Once we have the clusters of NXDOMAINs, we extract a feature vector for
    each cluster, which will be used for classification. We have four feature
    families: length, entropy, pairwise jaccard distance of character
    distribution, and pairwise dice distance of bigram distribution. This
    yields a 36-length feature vector for classification that relies on
    properties of the domain strings themselves. Please refer to Section 4.1.1
    in the original Pleiades paper~\cite{pleiades} for further details.

  \item Finally, the classifier uses the feature vectors of the clusters to
    detect existing, known DGAs and identify never-before-seen DGAs.

\end{enumerate}

To obtain DGA domains as a training dataset for the classifier, we analyzed
dynamic malware execution traffic and executed reverse-engineered DGA
algorithms.  Firstly, we identified NXDOMAINs that were queried by malware md5s
by analyzing malware pcaps obtained from a security vendor.  We used
AVClass~\cite{sebastian2016avclass} to get the malware family labels of those
md5s. Using this method, we labeled pykspa, suppobox, and gimemo malware
families, which were active in our dataset.
We extract one feature family per cluster for these.
Secondly, we use reverse engineered DGA domains to compensate
limited visibility of DGAs active in the network dataset.
Although only Pykspa, Suppobox, and Murofet domains have matches
in active clusters, we extract one feature vector for each version's daily domains
of 14 DGA families from the reversed engineered DGA domain dataset.
Table~\ref{tab:dga-families} shows the distribution of
the number of features vectors from the reverse engineered DGAs (top) and
those seen in clusters (bottom).

We trained the classifier with 17 classes, including 16 malware families
and one manually labeled benign class. We labeled
benign class from clusters containing mixture of all kinds of benign domains, as
well as clusters containing disposable domains (e.g., DNS queries to Anti-Virus
online reputation products~\cite{dnsfair}).

\begin{figure}
    \centering
    \scalebox{0.35}{\includegraphics{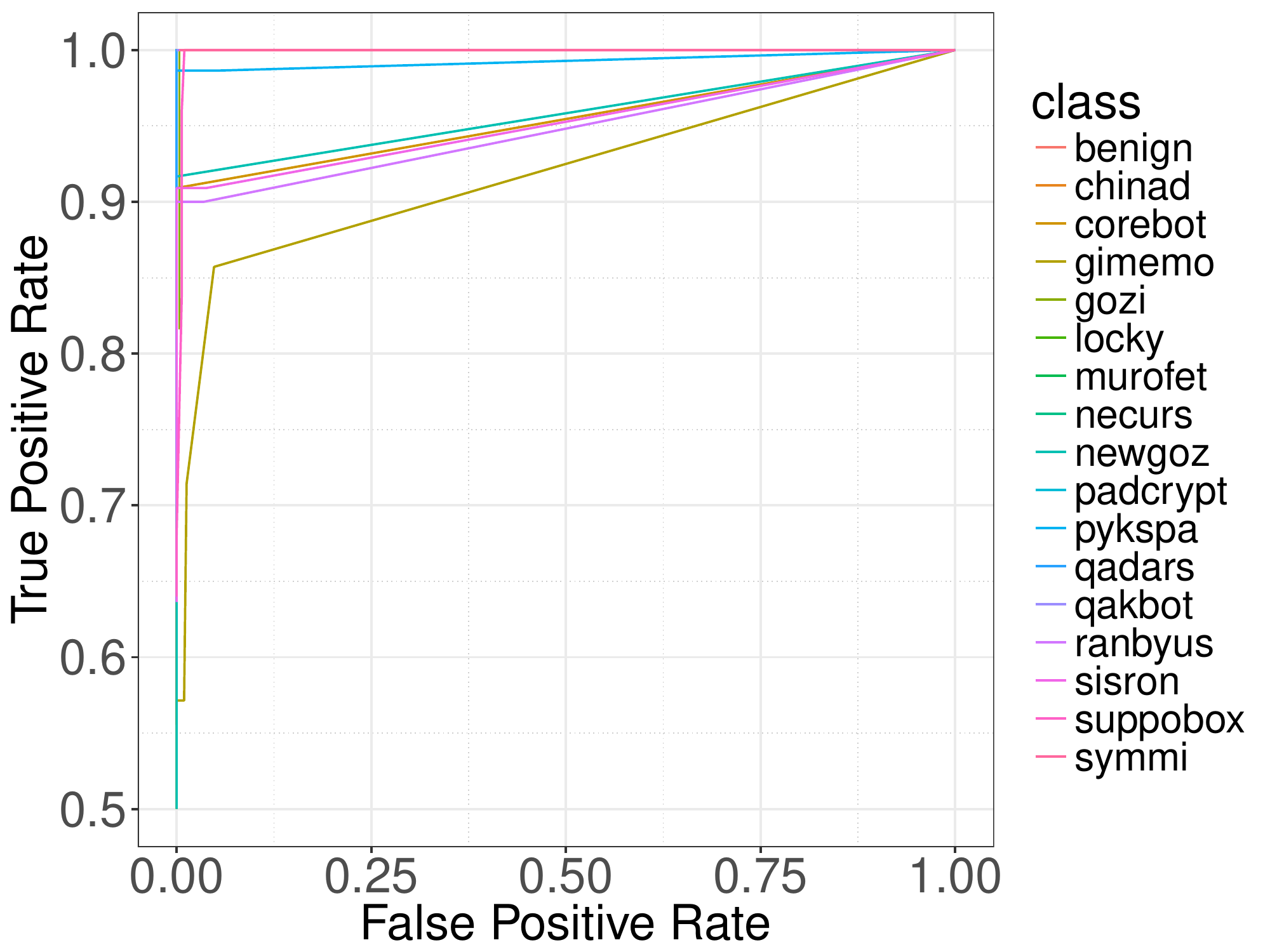}}
    \caption{
    ROC curves for 16 malware DGA classes and one benign class.
    }
   \label{fig:multiclassroc}
\end{figure}

\begin{figure}
    \centering
    \scalebox{0.3}{\includegraphics{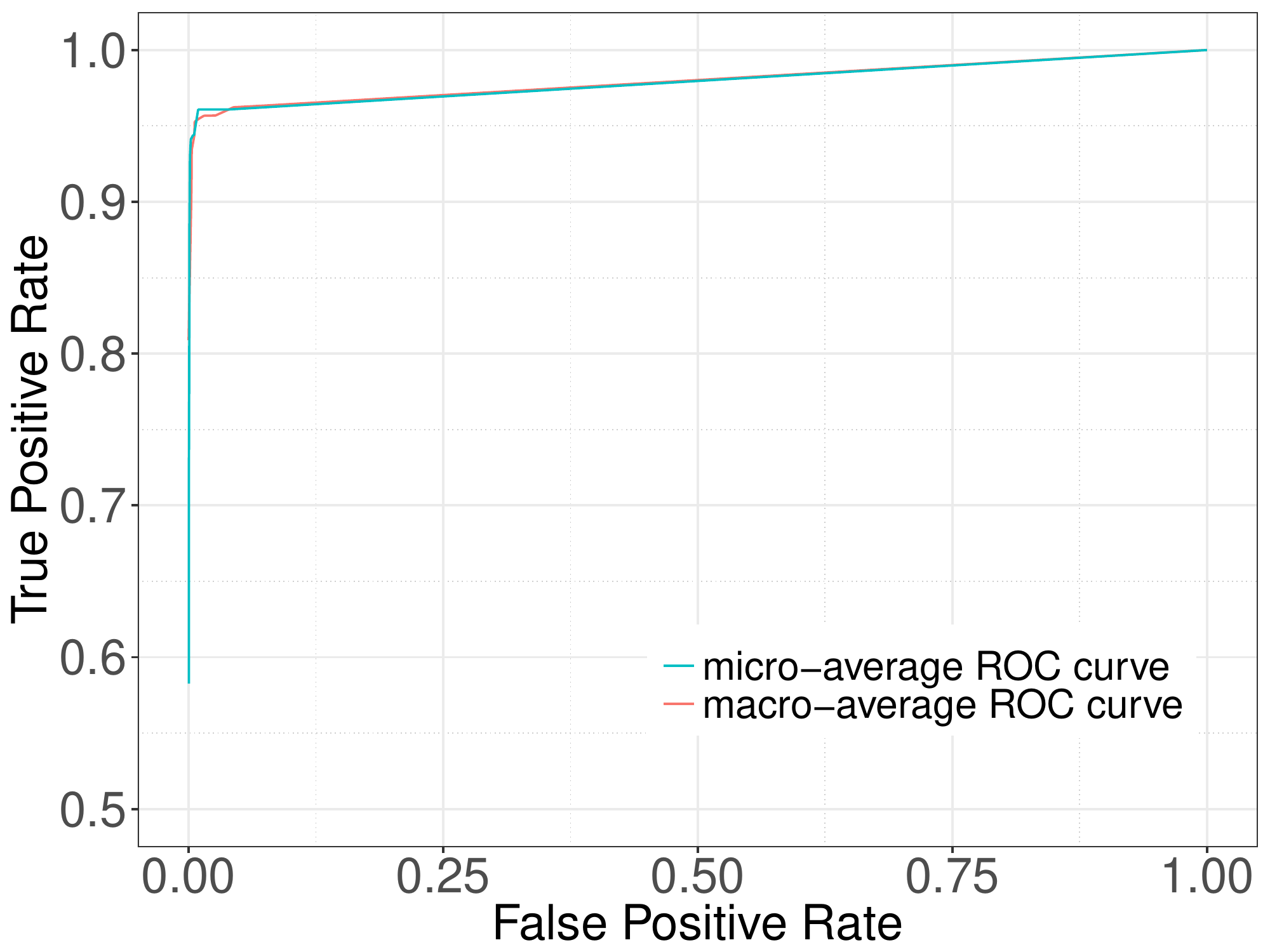}}
    \caption{
    Micro and macro ROC curves.
    }
   \label{fig:roc}
\end{figure}

We performed model selection to choose among the following algorithms:
Naive Bayes, Linear SVM, Random Forest, Logistic Regression and Stochastic
Gradient Descent Classifier. After the analysis of the performance of the
different classifiers, we chose to use Random Forest as our classifier.
Random Forests are similar to Alternative Decision Trees, a boosted
tree-based classifier, which were used in the original Pleiades paper.  We
tested our classifier with five fold cross-validation and measured an
average accuracy at \accuracy{}, and a false positive rate of \fpr{}.
Figure~\ref{fig:multiclassroc}
shows the multi-class ROC curves of the classifier performance.
Figure~\ref{fig:roc} shows the micro and macro ROC curves of the
multi-class classifier in our implementation of Pleiades.

\subsection{Current DGA Landscape} 
\label{sub:Current DGA Landscape}

We ran the DGA detection system over anonymized network traffic from a
Recursive DNS server in a telecommunication  provider, from December 18, 2016 to
December 29, 2016. 

\begin{figure}
    \centering
    \scalebox{0.43}{\includegraphics{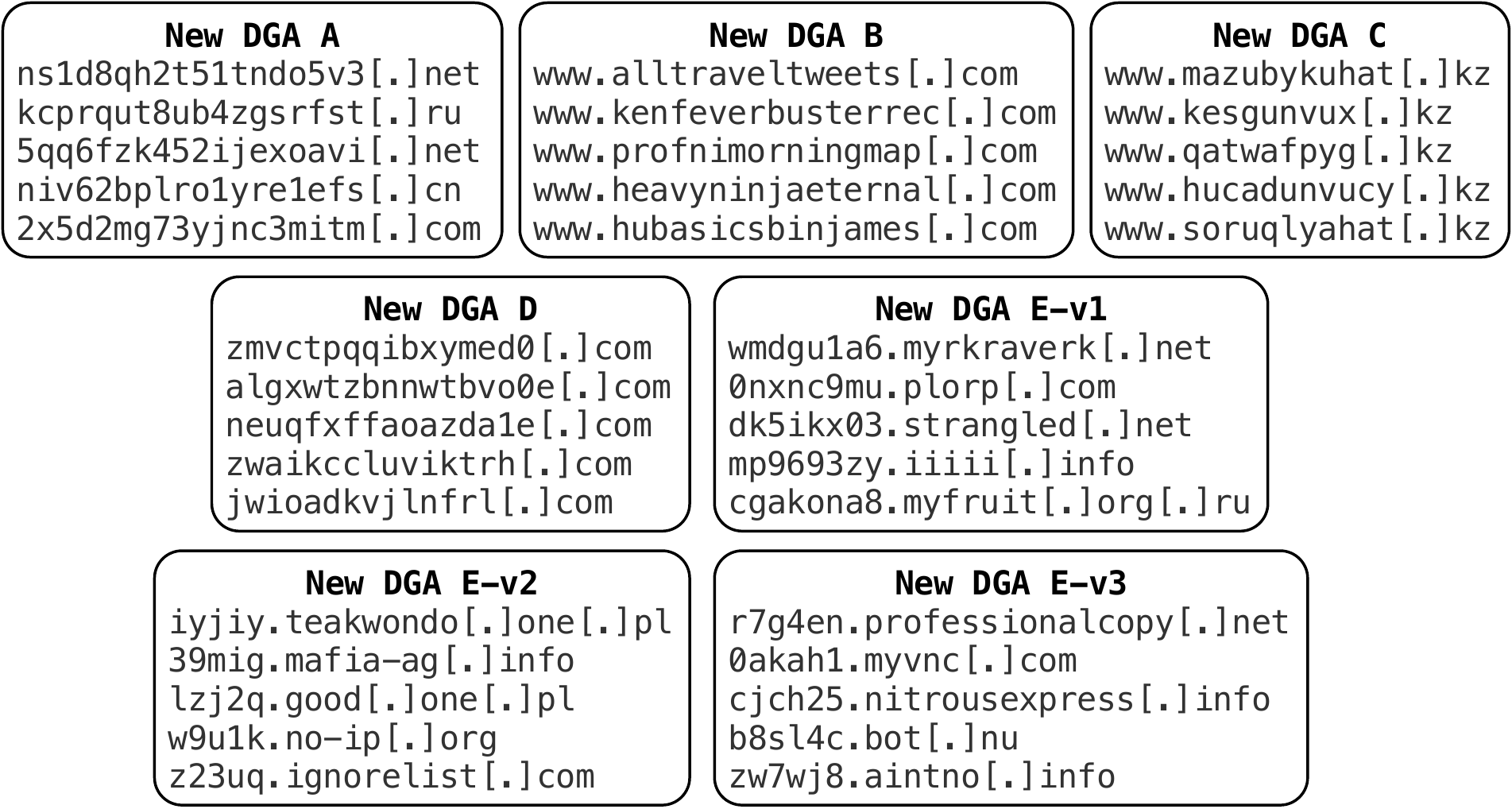}}
    \caption{
    Newly found DGAs.
    }
   \label{fig:newdgas}
\end{figure}


\paragraph{Newly Discovered DGAs} 
\label{par:Newly Discovered DGAs}

We found 12 new DGA malware families. Figure~\ref{fig:newdgas} shows 5 of them.
New DGA A is classified as similar to the DGA Chinad, with a
total of 59,904 domains.  The generated domains have a fixed length of 18
characters and use five different tlds: .com, .net, .cn, .biz, and .ru.  Chinad
has similar characteristics in domain names, but its domain length is 16
characters, and it uses two additional tlds: .info and .org.
New DGA B is a dictionary-words DGA that is classified as similar to Gozi.  Gozi
generates domains by combining words from word lists such as Requests for
Comments (RFC), the Ninety Five Theses of Martin Luther in its original Latin
text, and GNU General Public License (GNU GPL). In 12 days, we observed
9815 domain names from this DGA, with 10,435 infected hosts.
New DGA C is classified as similar to Gimemo. It repeatedly uses bigrams and
trigrams as units for composing domain name strings.  We found 6,738 domains
for new DGA C. Most of the domains from DGA C follow a pattern of
consonant-vowel-consonant at the beginning, usually followed by another similar
pattern or a sequence of vowel-consonant-vowel, which makes the generated
domains appear almost readable. Nevertheless, New DGA C generated domains did not
follow the character frequency distribution for any of the languages that use
the English or similar alphabets.  The length of the generated domains is not
fixed but it appears to be around 10 characters with either a character added
or removed.
New DGA D uses \texttt{.com} tld, and second-level labels varying between 12 and 18 characters.
New DGA E-v1 iterates through both algorithm-generated
second level domains and child labels.


\paragraph{Evasion Attempts in the Wild} 
\label{par:Evasion Attempts in the Wild}

The DGAs of qakbot and pykspa provide us with evidence that the malware authors
are attempting to avoid or obstruct detection.  A special mode of Qakbot is
triggered when the malware detects that it is running inside a sandbox environment.
Specifically, the seed of the algorithm is appended to generate redundant
domains that won't be used as actual C\&C. Similary, Pykspa generates two sets
of domains based on two different seed sets, which appear identical to a human
analyst as if there were only one set of generated domains.  Different than
Qakbot, in normal operation Pykspa queries both sets of domains, along a list
of benign domains. This kind of behavior could be a method to detect analysis
efforts. If an analyst sets the environment to provide answers to these
``bogus'' queries, it could indicate anomaly to the malware.  Generating a large
number of ``fake'' domains could also increase the cost of sinkholing the
botnets.  It makes the sinkholing operation more likely to fail to cover all of
the actual C\&C domains~\cite{nadji2013beheading}.  These efforts appear to be
in their infancy in terms of complexity and effectiveness at this point. If
malware authors unleash their creativity in the future, we might come across
more elaborate evasion cases that require a lot more effort to identify and
detect.

Furthermore, we identify instances of DGAs already evading the classification
part of Pleiades by introducing a child label. Our classifier has low
confidence for detecting new DGAs B, C, and E-v1.
Since there are no DGA domains with child labels in the training dataset,
the classifier does not have the requisite knowledge to predict such DGAs.
After deploying the classifier for 12 days, we retrained the classifier
with additional DGA families observed from the network.
After retraining, our classifier has successfully identified the following new variants
with high confidence: DGA E-v2 and DGA E-v3.

\end{document}